\documentclass[aps,prd,twocolumn,groupedaddress,10pt]{revtex4-1}
\usepackage{amssymb}
\usepackage{graphicx}
\usepackage{amsmath}
\usepackage{hyperref}

\begin{document}

\title{Reheating phase diagram for single-field slow-roll inflationary models}

\author{Rong-Gen Cai}
\email{cairg@itp.ac.cn}
\author{Zong-Kuan Guo}
\email{guozk@itp.ac.cn}
\author{Shao-Jiang Wang}
\email{schwang@itp.ac.cn}
\affiliation{State Key Laboratory of Theoretical Physics, Institute of Theoretical Physics, Chinese Academy of Sciences, Beijing 100190, China}
\date{\today}
\begin{abstract}
  We investigate the influence on the inflationary predictions from the reheating processes characterized by the $e$-folding number $N_{\mathrm{reh}}$ and the effective equation-of-state parameter $w_{\mathrm{reh}}$ during the reheating phase. For the first time, reheating processes can be constrained in the $N_{\mathrm{reh}}\!-\!w_{\mathrm{reh}}$ plane from Planck 2015. We find that for Higgs inflation with a nonminimal coupling to gravity, the predictions are insensitive to the reheating phase for current CMB measurements. We also find that the spontaneously broken SUSY inflation and axion monodromy inflation with $\phi^{2/3}$ potential, which with instantaneous reheating lie outside or at the edge of the $95\%$ confidence region in the $n_s\!-\!r$ plane from Planck 2015 TT,TE,EE$+$lowP, can well fit the data with the help of reheating processes. Future CMB experiments would put strong constraints on reheating processes.
\end{abstract}
\maketitle

\section{Introduction}\label{sec:1}

Models of single-field slow-roll inflation with a standard kinetic term provide a very good fit to the Planck data~\cite{Ade:2013uln,Ade:2015lrj}. In the inflationary scenario, the Universe expands quasi-exponentially as the scalar field rolls slowly along a very flat potential.
After inflation ends, the inflaton field oscillates around the minimum of the potential.
During such a period of reheating, the energy in the inflaton is transferred to the plasma of standard model particles.
Reheating is an integral part of inflationary models. Without reheating, the Universe after inflation would be empty and cold.
However, the physics of reheating may be more complicated. As pointed out in Ref.~\cite{Podolsky:2005bw}, the equation-of-state parameter changes sharply during the reheating phase due to the out-of-equilibrium nonlinear dynamics of fields.

Predictions of slow-roll inflationary models are usually derived given a reasonable range for the $e$-folding number $N_{\mathrm{inf}}$ during inflation. With more and more precise CMB measurements~\cite{Ade:2013uln,Ade:2015lrj}, it becomes important to consider the impact on the inflationary predictions from reheating processes~\cite{Martin:2010kz,Martin:2010hh,Martin:2013tda,Martin:2013nzq,Martin:2014rqa,Martin:2014nya}. It is pointed out that the reheating processes provide additional constraints on inflationary models via the reheating temperature~\cite{Dai:2014jja}, and some specific inflationary models are investigated in Refs.~\cite{Creminelli:2014fca,Munoz:2014eqa,Cook:2015vqa}. Although the physics of reheating is highly uncertain and unconstrained due to its nonlinear backreaction and nonperturbative nature, the reheating phase can in principle be characterized by two parameters only: the $e$-folding number $N_{\mathrm{reh}}$ and the effective equation-of-state (EoS) parameter $w_{\mathrm{reh}}$. In terms of these two parameters, one can express the observable quantities such as the scalar spectral index $n_s$ and its running $\alpha_s$, the tensor spectral index $n_t$ and its running $\alpha_t$, the tensor-to-scalar ratio $r$, the $e$-folding number $N_{\mathrm{inf}}$, and the reheating temperature $T_{\mathrm{reh}}$. Thus, reheating processes can be constrained in the $N_{\mathrm{reh}}\!-\!w_{\mathrm{reh}}$ plane by Planck constraints on $n_s$ and $r$.

Following the approach proposed in Refs.~\cite{Dai:2014jja,Creminelli:2014fca,Munoz:2014eqa}, in this paper we study several models of single-field slow-roll inflation including Higgs inflation, power-law potential, hilltop inflation, natural inflation, spontaneously broken (SB) SUSY inflation, and superconformal $\alpha$-attractors. Although the reheating/preheating mechanisms are well studied in the Higgs inflation~\cite{Bezrukov:2007ep,Bezrukov:2008ut,GarciaBellido:2008ab}, the underlying mechanisms of unitarization and stabilization in Higgs inflation might have substantial impact on the reheating phase, as discussed in Ref.~\cite{Bezrukov:2014ipa}. For Higgs inflation, we find that the inflationary predictions are insensitive to reheating processes given the current precision of CMB measurements. The same conclusion is also derived in Ref.~\cite{Gong:2015qha}.
We find that the SB SUSY inflation and power-law potential $\phi^{2/3}$, which lie outside or at the edge of the $95\%$ confidence region in the $n_s\!-\!r$ plane from Planck 2015 TT,TE,EE$+$lowP, can well fit the data if reheating processes are taken into account. However the constrained parameter space of reheating processes is still very large for most inflationary models due to current relatively weak constraints on inflation; future measurements of $n_s$ and $r$ will eventually narrow down the parameter space, thus revealing the physics of the reheating era.

The paper is organized as follows: In Sec.~\ref{sec:2} we introduce the effective descriptions of the reheating phase in terms of $N_{\mathrm{reh}}$ and $w_{\mathrm{reh}}$ for single-field slow-roll inflationary models. In Sec.~\ref{sec:3} we study some specific inflationary models. Section~\ref{sec:4} is devoted to conclusions.

\section{Descriptions of the Reheating Phase}\label{sec:2}

First, following the method proposed in Refs.~\cite{Dai:2014jja, Creminelli:2014fca, Munoz:2014eqa}, we derive a formula for the effective number of degrees of freedom $g_{\mathrm{reh}}$ at the end of the reheating phase. It is worth noting that one should not take this literally, since most of the observables are insensitive to the precise value of $g_{\mathrm{reh}}$ due to its logarithmic dependence. However, the derived formula of $g_{\mathrm{reh}}$ will be essential to carry out the inflationary predictions in the $N_{\mathrm{reh}}\!-\!w_{\mathrm{reh}}$ plane, which can be used to constrain the parameter space of the reheating phase to meet the current constraints on inflation.

The pivot scale is chosen as $k_*=0.05\,\mathrm{Mpc}^{-1}$, which is also expressed by
\begin{equation}\label{eq:k=aH}
k_*=a_*H_*=\frac{a_*}{a_{\mathrm{end}}}\frac{a_{\mathrm{end}}}{a_{\mathrm{reh}}}\frac{a_{\mathrm{reh}}}{a_0} a_0 H_*.
\end{equation}
In what follows, the current scale factor $a_0=1$ and all quantities with subscript ``$*$" are evaluated at the moment when the pivot scale crosses the horizon.

The first two factors of (\ref{eq:k=aH}) can be computed by
\begin{equation}\label{eq:part1}
\frac{a_*}{a_{\mathrm{end}}}\frac{a_{\mathrm{end}}}{a_{\mathrm{reh}}}=e^{-(N_*+N_{\mathrm{reh}})}.
\end{equation}

The third factor of (\ref{eq:k=aH}) is computed as follows. We use the conservation equation of entropy $g_{\mathrm{reh}}a_{\mathrm{reh}}^3T_{\mathrm{reh}}^3=g_{\gamma}a_0^3T_{\gamma}^3+g_{\nu}a_0^3T_{\nu}^3=(43/11)T_{\gamma}^3a_0^3$ by noting that
$g_{\gamma}=2$, $g_{\nu}=(7/8)\times3\times2=21/4$, and $T_{\nu}^3=(4/11)T_{\gamma}^3$; here we adopt $T_{\gamma}=2.7255$ K.
Thus, the third factor can be written as
\begin{equation}\label{eq:part2}
\frac{a_{\mathrm{reh}}}{a_0}=\left(\frac{43}{11g_{\mathrm{reh}}}\right)^{\frac{1}{3}}\frac{T_{\gamma}}{T_{\mathrm{reh}}},
\end{equation}
where the temperature $T_{\mathrm{reh}}$ at the end of the reheating phase needs to be properly accounted for. Assuming that the whole history during the reheating phase can be effectively described by the $e$-folding number $N_{\mathrm{reh}}$ and the effective EoS parameter $w_{\mathrm{reh}}$, we are able to relate the reheating phase to the inflationary phase via
\begin{align}
 \rho_{\mathrm{end}}&=\frac{3}{3-\epsilon_{\mathrm{end}}}V_{\mathrm{end}},\\
 \rho_{\mathrm{reh}} &=\rho_{\mathrm{end}}\,e^{-3N_{\mathrm{reh}}(1+w_{\mathrm{reh}})},\\
 \rho_{\mathrm{reh}} &=\frac{\pi^2}{30}g_{\mathrm{reh}}T_{\mathrm{reh}}^4.
\end{align}
The inflation ends when the first Hubble hierarchy parameter $\epsilon_H=(3/2)(1+w)=3K/(K+V)=\epsilon_{\mathrm{end}}$; therefore the kinetic energy $K_{\mathrm{end}}=\epsilon_{\mathrm{end}}V_{\mathrm{end}}/(3-\epsilon_{\mathrm{end}})$; thus the total energy density $\rho_{\mathrm{end}}=K_{\mathrm{end}}+V_{\mathrm{end}}=3V_{\mathrm{end}}/(3-\epsilon_{\mathrm{end}})$. The second equation above comes from direct calculation of $\rho_{\mathrm{reh}}=\rho_{\mathrm{end}}\exp\left(-3\int_{a_{\mathrm{end}}}^{a_{\mathrm{reh}}}(1+w_{\mathrm{reh}})\mathrm{d}\ln a\right)=\rho_{\mathrm{end}}\,e^{-3N_{\mathrm{reh}}(1+w_{\mathrm{reh}})}$. Combining the above three equations gives rise to
\begin{equation}\label{eq:Treh}
T_{\mathrm{reh}}^4=\frac{90V_{\mathrm{end}}}{\pi^2(3-\epsilon_{\mathrm{end}})g_{\mathrm{reh}}}\,e^{-3N_{\mathrm{reh}}(1+w_{\mathrm{reh}})}.
\end{equation}
Inserting (\ref{eq:Treh}) into (\ref{eq:part2}) gives the final expression for the third factor of (\ref{eq:k=aH}).

The last factor $H_*$ is usually fixed by Planck normalization $A_s=H_*^2/8\pi^2\epsilon_*$, namely $H_*=\pi\sqrt{r_*A_s/2}$. However, it requires full knowledge on the tensor-to-scalar ratio $r_*$ and the consistency relation $r_*=16\epsilon_*$, both of which have not been observed or confirmed yet. A more conservative way to compute $H_*$ is to use the slow-roll equation
\begin{equation}\label{eq:part3}
 3H_*^2=V_*.
\end{equation}

Combining (\ref{eq:part1}), (\ref{eq:part2}), and (\ref{eq:part3}) together gives the final formula for the effective number of degrees of freedom at the end of the reheating phase:
\begin{align}\label{eq:greh1}
 \nonumber g_{\mathrm{reh}}&=\left(\frac{T_{\gamma}}{k_*}\right)^{12}\left(\frac{43}{11}\right)^4\left(\frac{3-\epsilon_{\mathrm{end}}}{810}\right)^3\left(\frac{\pi^6V_*^6}{V_{\mathrm{end}}^3}\right)\\
                 &\times\exp\left(9N_{\mathrm{reh}}\left(w_{\mathrm{reh}}-\frac{1}{3}\right)-12N_*\right).
\end{align}

Next, we apply (\ref{eq:greh1}) for general single-field slow-roll inflationary potential $V(\phi,p)$ with only one parameter $p$ (for multiparameter inflationary potential, one has to fix some of its parameters).  The inflation ended when the slow-roll condition was broken:
\begin{equation}\label{eq:phiend}
\epsilon(\phi_{\mathrm{end}},p)=\epsilon_{\mathrm{end}}\Rightarrow\phi_{\mathrm{end}}(p).
\end{equation}
Once we have the field value of $\phi_{\mathrm{end}}(p)$ at the end point of inflation, we have the potential energy density at that moment:
\begin{equation}\label{eq:Vend}
V(\phi_{\mathrm{end}}(p),p)\equiv V_{\mathrm{end}}(p).
\end{equation}
To compute $H_*$ via the slow-roll equation $3H_*^2=V_*$, one requires the field value of $\phi_*$ when the pivot scale crosses the horizon. This can only be done by inputting Planck observations on the scalar power spectrum amplitude via
\begin{equation}\label{eq:phi*}
A_s=\frac{1}{24\pi^2}\frac{V(\phi_*,p)}{\epsilon(\phi_*,p)}\Rightarrow\phi_*(p).
\end{equation}
Once we know the field value of $\phi_*(p)$, we have all the information needed to restore $H_*=\sqrt{V_*/3}$ by
\begin{equation}\label{eq:V*}
V(\phi_*(p),p)\equiv V_*(p).
\end{equation}
There is one more quantity to complete the evaluations, which is the $e$-folding number during inflation:
\begin{equation}\label{eq:N*}
\int_{\phi_{\mathrm{end}}(p)}^{\phi_*(p)}\frac{V(\phi,p)}{V'_{\phi}(\phi,p)}\mathrm{d}\phi\equiv N_*(p).
\end{equation}
Combining (\ref{eq:Vend}), (\ref{eq:V*}), and (\ref{eq:N*}) together, we have
\begin{align}
 \nonumber g_{\mathrm{reh}}&(p,N_{\mathrm{reh}},w_{\mathrm{reh}})\\
 \nonumber &=\left(\frac{T_{\gamma}}{k_*}\right)^{12}\left(\frac{43}{11}\right)^4
             \left(\frac{3-\epsilon_{\mathrm{end}}}{810}\right)^3\,\frac{\pi^6V_*^6(p)}{V_{\mathrm{end}}^3(p)}\\
           &\times\exp\left(9N_{\mathrm{reh}}\left(w_{\mathrm{reh}}-\frac{1}{3}\right)-12N_*(p)\right)\label{eq:greh2}.
\end{align}

Finally, three comments on (\ref{eq:greh2}) follow:
\begin{enumerate}
  \item It can be tested that the general formalism presented above is insensitive to the precise values of $\epsilon_{\mathrm{end}}$ and $g_{\mathrm{reh}}$ due to the logarithmic dependence. Therefore, it suffices to take the fiducial values $\epsilon_{\mathrm{end}}=1$,  $g_{\mathrm{reh}}=106.75$ for Higgs inflation and $g_{\mathrm{reh}}=10^3$ for other single-field slow-roll inflationary models.
  \item The $w_{\mathrm{reh}}=1/3$ case should be seen as the equivalent case $N_{\mathrm{reh}}=0$, which presents an instantaneous reheating process. We will see in the next section that an instantaneous reheating process manifests itself as an asymptotic line in the $N_{\mathrm{reh}}\!-\!w_{\mathrm{reh}}$ plane.
  \item The degeneracy between between the phases is the source of the freedom of choices of $N_*$ in the $n_s\!-\!r$ plane. This can be seen from the fact that any shift $\Delta N_*$ from $N_*$ can be compensated by the shifts $\Delta N_{\mathrm{reh}}$ and $\Delta w_{\mathrm{reh}}$ from $N_{\mathrm{reh}}$ and $w_{\mathrm{reh}}$, provided that $9N_{\mathrm{reh}}\Delta w_{\mathrm{reh}}+9\Delta N_{\mathrm{reh}}(w_{\mathrm{reh}}+\Delta w_{\mathrm{reh}}-\frac{1}{3})=12\Delta N_*$.
\end{enumerate}

\section{Reheating Phase Diagram}\label{sec:3}

We start with the usual $n_s\!\!-\!r$ plane. By solving the equations
\begin{align}
A_s&=\frac{1}{24\pi^2}\frac{V(\phi_*,p)}{\epsilon(\phi_*,p)},\\
N_*&=\int_{\phi_{\mathrm{end}}(p)}^{\phi_*}\frac{V(\phi,p)}{V'_{\phi}(\phi,p)}\mathrm{d}\phi
\end{align}
with an input value of $A_s$, one can express the field value $\phi_*(N_*)$ and potential parameter $p(N_*)$ in terms of $e$-folding number $N_*$ during inflation. Hence, the inflationary predictions $n_s(\phi_*(N_*),p(N_*))\equiv n_s(N_*)$ and $r(\phi_*(N_*),p(N_*))\equiv r(N_*)$ can be made with respect to $N_*$ in the $n_s\!-\!r$ plane. However, the reheating phase variables $N_{\mathrm{reh}}$ and $w_{\mathrm{reh}}$ cannot be specified in this case in the $n_s\!\!-\!r$ plane.

To break the degeneracy between the inflationary phase and the reheating phase as mentioned in the previous section, we propose to solve the equations
\begin{align}
A_s&=\frac{1}{24\pi^2}\frac{V(\phi_*,p)}{\epsilon(\phi_*,p)},\\
g_{\mathrm{reh}}&=g_{\mathrm{reh}}(p,N_{\mathrm{reh}},w_{\mathrm{reh}})\label{eq:greh3}
\end{align}
with input values of $A_s$ and $g_{\mathrm{reh}}$, and the obtained solutions $\phi_*(N_{\mathrm{reh}},w_{\mathrm{reh}})$ and $p(N_{\mathrm{reh}},w_{\mathrm{reh}})$ can be used to express inflationary observables like $n_s(\phi_*(N_{\mathrm{reh}},w_{\mathrm{reh}}),p(N_{\mathrm{reh}},w_{\mathrm{reh}}))\equiv n_s(N_{\mathrm{reh}},w_{\mathrm{reh}})$, $r(\phi_*(N_{\mathrm{reh}},w_{\mathrm{reh}}),p(N_{\mathrm{reh}},w_{\mathrm{reh}}))\equiv r(N_{\mathrm{reh}},w_{\mathrm{reh}})$ in terms of reheating phase variables $N_{\mathrm{reh}}$ and $w_{\mathrm{reh}}$. Other quantities like $N_*(p(N_{\mathrm{reh}},w_{\mathrm{reh}}))\equiv N_*(N_{\mathrm{reh}},w_{\mathrm{reh}})$ in (\ref{eq:N*}) and $T_{\mathrm{reh}}(p(N_{\mathrm{reh}},w_{\mathrm{reh}}),N_{\mathrm{reh}},w_{\mathrm{reh}})\equiv T_{\mathrm{reh}}(N_{\mathrm{reh}},w_{\mathrm{reh}})$ in (\ref{eq:Treh}) can also be expressed in terms of reheating phase variables $N_{\mathrm{reh}}$ and $w_{\mathrm{reh}}$. Expressing various observables in terms of reheating phase variables $N_{\mathrm{reh}}$ and $w_{\mathrm{reh}}$ in the $N_{\mathrm{reh}}\!-\!w_{\mathrm{reh}}$ plane will be referred as the \emph{reheating phase diagrams}. It can be tested that the reheating phase diagrams are insensitive to different input values of $g_{\mathrm{reh}}$.

What priors should we choose for $N_{\mathrm{reh}}$ and $w_{\mathrm{reh}}$ in general? First, the inflation era ends when the EoS parameter equals $-1/3$, and the radiation era begins when the EoS parameter equals $1/3$. It seems that $w_{\mathrm{reh}}$ should be in the range $[-1/3,1/3]$. However, it is possible to achieve potential dominance (with its EoS parameter equal to $-1$) and kinetic dominance (with its EoS parameter equal to $1$), assuming a massive inflaton. Second, in the $n_s\!-\!r$ plane, the inflationary predictions are usually made by choosing $N_{\mathrm{inf}}$ in the range $[50,60]$, or more generally $[40,70]$, which is actually degenerated with $N_{\mathrm{reh}}$ and $w_{\mathrm{reh}}$, because, as mentioned in the last section, any shift $\Delta N_*$ from $N_*$ can be compensated by the shifts $\Delta N_{\mathrm{reh}}$ and $\Delta w_{\mathrm{reh}}$ from $N_{\mathrm{reh}}$ and $w_{\mathrm{reh}}$, provided that $9N_{\mathrm{reh}}\Delta w_{\mathrm{reh}}+9\Delta N_{\mathrm{reh}}(w_{\mathrm{reh}}+\Delta w_{\mathrm{reh}}-\frac{1}{3})=12\Delta N_*$. Assuming the maximum $e$-folding number during inflation $N_*=70$ which can be shifted by $\Delta N_*=-30$, and the maximum EoS parameter during reheating $w_{\mathrm{reh}}=1$ which can be shifted by $\Delta w_{\mathrm{reh}}=-2$, one can easily work out the minimum $e$-folding number during reheating $N_{\mathrm{reh}}=0$, which can be shifted by
\begin{equation}
\Delta N_{\mathrm{reh}}=\frac{12\Delta N_*-9N_{\mathrm{reh}}\Delta w_{\mathrm{reh}}}{9\left(w_{\mathrm{reh}}+\Delta w_{\mathrm{reh}}-\frac{1}{3}\right)}=30.
\end{equation}
Therefore, without preknowledge on the reheating phase, one can in general choose $N_{\mathrm{reh}}$ in the prior $[0,30]$ and $w_{\mathrm{reh}}$ in the prior $[-1,1]$, which will cover $N_*$ in the range $[40,70]$. A constant EoS parameter $w_{\mathrm{reh}}$ in this sense should be viewed as an effective parameter time-averaging the EoS parameter during the whole reheating process.

\subsection{Reheating phase diagram for Higgs inflation}\label{subsec:3.1}

In the Higgs inflation, the Higgs field with a large nonminimal coupling to Einstein gravity in the Jordan frame can give rise to an exponential plateau-like potential in the large field region in the Einstein frame where the inflaton is defined. The action in the Jordan frame is
\begin{equation}\label{eq:Jordan Frame}
S_J=\int\mathrm{d}^4x\sqrt{-g}\left(\frac{M_{\mathrm{P}}^2}{2}\Omega^2R-\frac{1}{2}(\partial h)^2-V(h)\right),
\end{equation}
where $V(h)=(\lambda/4)(h^2-v^2)^2$ with the vacuum expectation value (VEV) of electroweak (EW) vacuum $v=246$ GeV. Here the conformal factor
\begin{equation}\label{eq:conformal factor}
\Omega^2=\frac{\widetilde{g}_{\mu\nu}}{g_{\mu\nu}}=1+\frac{\xi h^2}{M_{\mathrm{P}}^2}
\end{equation}
and the scalar field redefinition
\begin{equation}\label{eq:normalization}
\left(\frac{\mathrm{d}\chi}{\mathrm{d}h}\right)^2
=\frac{1}{\Omega^2}+\frac{6M_{\mathrm{P}}^2}{\Omega^2}\left(\frac{\mathrm{d}\Omega}{\mathrm{d}h}\right)^2
 \end{equation}
allow us to switch the action into the Einstein frame
\begin{equation}
S_E=\int\mathrm{d}^4x\sqrt{-\widetilde{g}}\left(\frac{M_{\mathrm{P}}^2}{2}\widetilde{R}-\frac{1}{2}(\widetilde{\partial}\chi)^2-U(\chi)\right),
\end{equation}
where the kinetic terms for Einstein gravity and the new scalar field are both canonically normalized. The potential term
\begin{equation}\label{eq:U(chi)}
U(\chi(h))=\frac{V(h)}{\Omega^4}=\frac{\lambda M_{\mathrm{P}}^4}{4\xi^2}\frac{\left(\frac{\xi h^2}{M_{\mathrm{P}}^2}-\frac{\xi v^2}{M_{\mathrm{P}}^2}\right)^2}{\left(1+\frac{\xi h^2}{M_{\mathrm{P}}}\right)^2}
\end{equation}
can be abbreviated as
\begin{equation}\label{eq:U(phi)}
U(\chi(\phi))=\frac{Z}{4}\frac{\phi^4}{(1+\phi^2)^2}\equiv V(\phi)
\end{equation}
by using the dimensionless scalar field $\phi=\sqrt{\xi}h/M_{\mathrm{P}}$ and the combined parameter $Z=\lambda/\xi^2$ for later convenience. Here we ignore $v$ and set $M_{\mathrm{P}}^2=1$ from now on. In the large field region $h\gg M_{\mathrm{P}}/\sqrt{\xi}$, one can solve the scalar field redefinition (\ref{eq:normalization}) to obtain $\chi\simeq\sqrt{6}M_{\mathrm{P}}\ln\sqrt{\xi}h/M_{\mathrm{P}}\equiv\sqrt{6}M_{\mathrm{P}}\ln\phi$; thus an exponential plateau-like potential in the large field region
\begin{equation}
U(\chi)=\frac{Z}{4}\left(1+e^{-\frac{2\chi}{\sqrt{6}}}\right)^{-2}
\end{equation}
is obtained as promised.
The slow-roll dynamics with respect to the inflaton $\chi$ can be carried out directly by computing the slow-roll parameters
\begin{eqnarray}
\epsilon(\phi)&=&\frac{4}{3}\frac{1}{(\phi^2+1)^2},\\
\eta(\phi)    &=&-\frac{4}{3}\frac{1}{\phi^2+1}+\frac{4}{(\phi^2+1)^2},\\
\zeta^2(\phi) &=&\frac{16/9}{(\phi^2+1)^2}-\frac{16}{(\phi^2+1)^3}+\frac{64/3}{(\phi^2+1)^4},
\end{eqnarray}
the $e$-folding number during inflation
\begin{equation}
N(\phi_N)=\frac{3}{4}\left(\phi_N^2-\phi_{\mathrm{end}}^2-\ln\frac{1+\phi_N^2}{1+\phi_{\mathrm{end}}^2}\right),
\end{equation}
the scalar spectral indexes and its running and the tensor-to-scalar ratio
\begin{align}
&n_s(\phi_N)=1-\frac{8}{3}\frac{1}{\phi_N^2+1},\\
&r(\phi_N)=\frac{64}{3}\frac{1}{(\phi_N^2+1)^2},\\
&\alpha_s(\phi_N)=-\frac{32}{9}\frac{1}{(\phi_N^2+1)^2}+\frac{32}{9}\frac{1}{(\phi_N^2+1)^3}
\end{align}
in terms of dimensionless $\phi$.

\begin{figure*}
  \includegraphics[width=7cm]{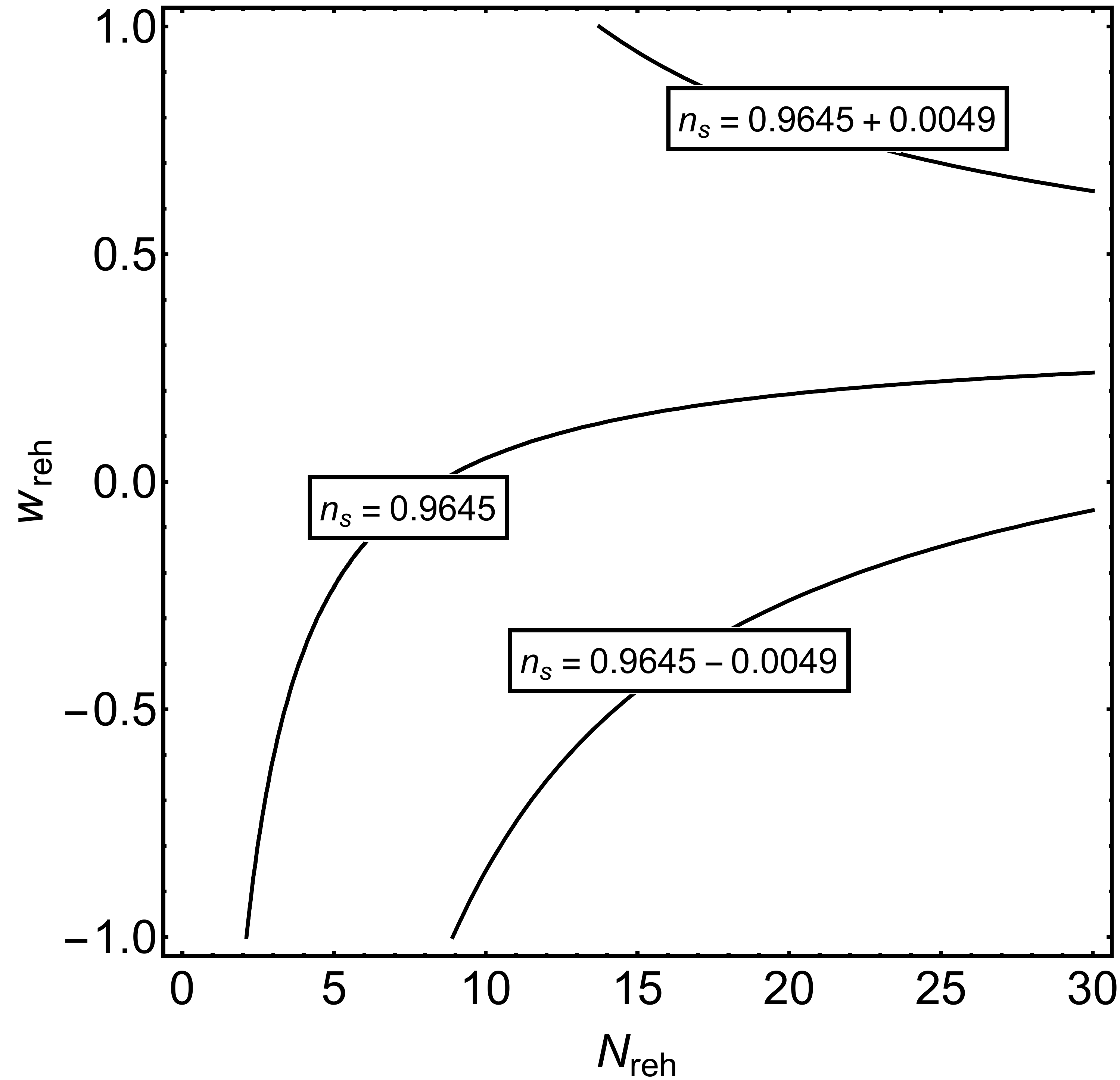}
  \includegraphics[width=7cm]{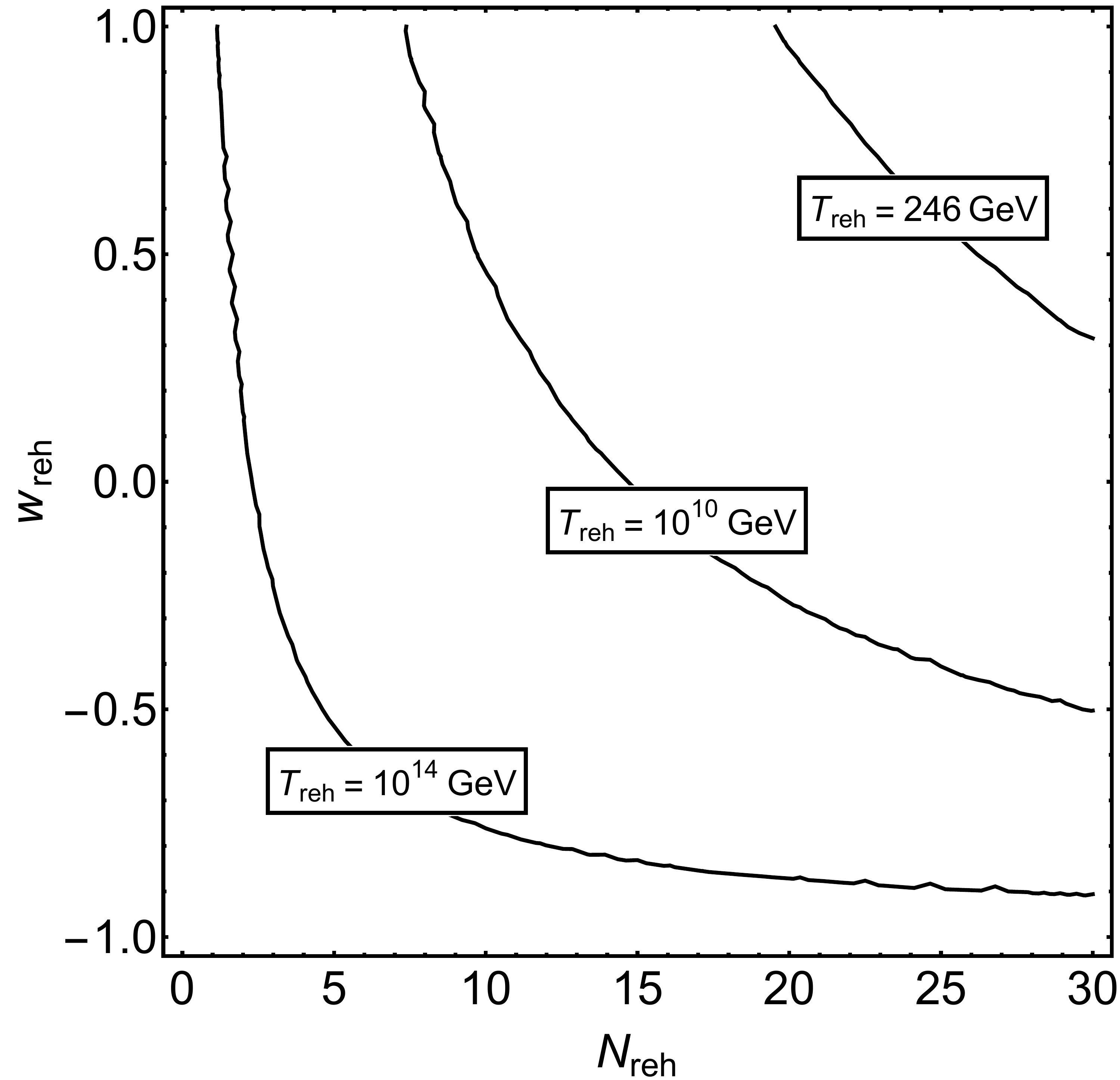}\\
  \includegraphics[width=7cm]{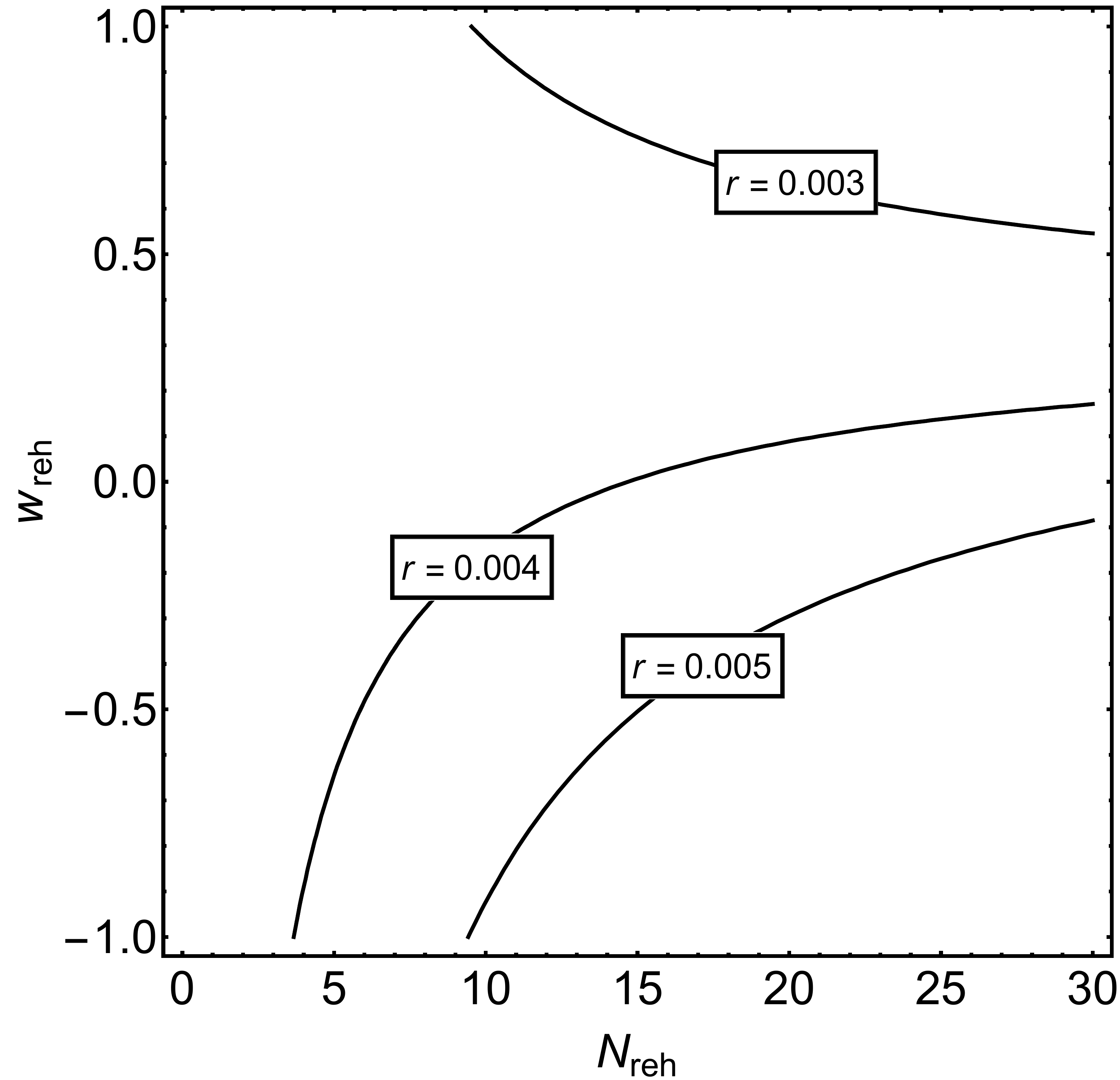}
  \includegraphics[width=7cm]{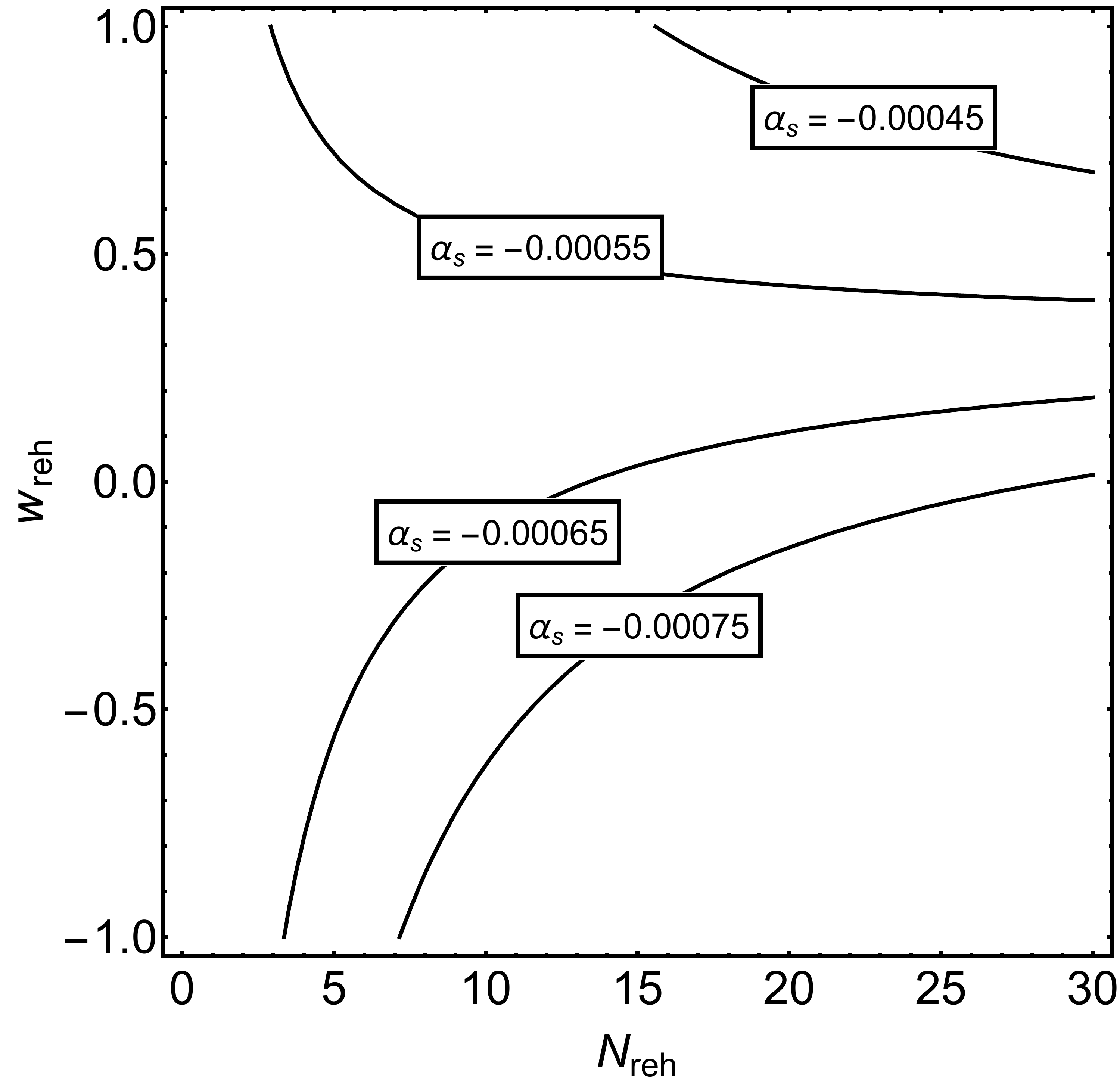}\\
  \includegraphics[width=7cm]{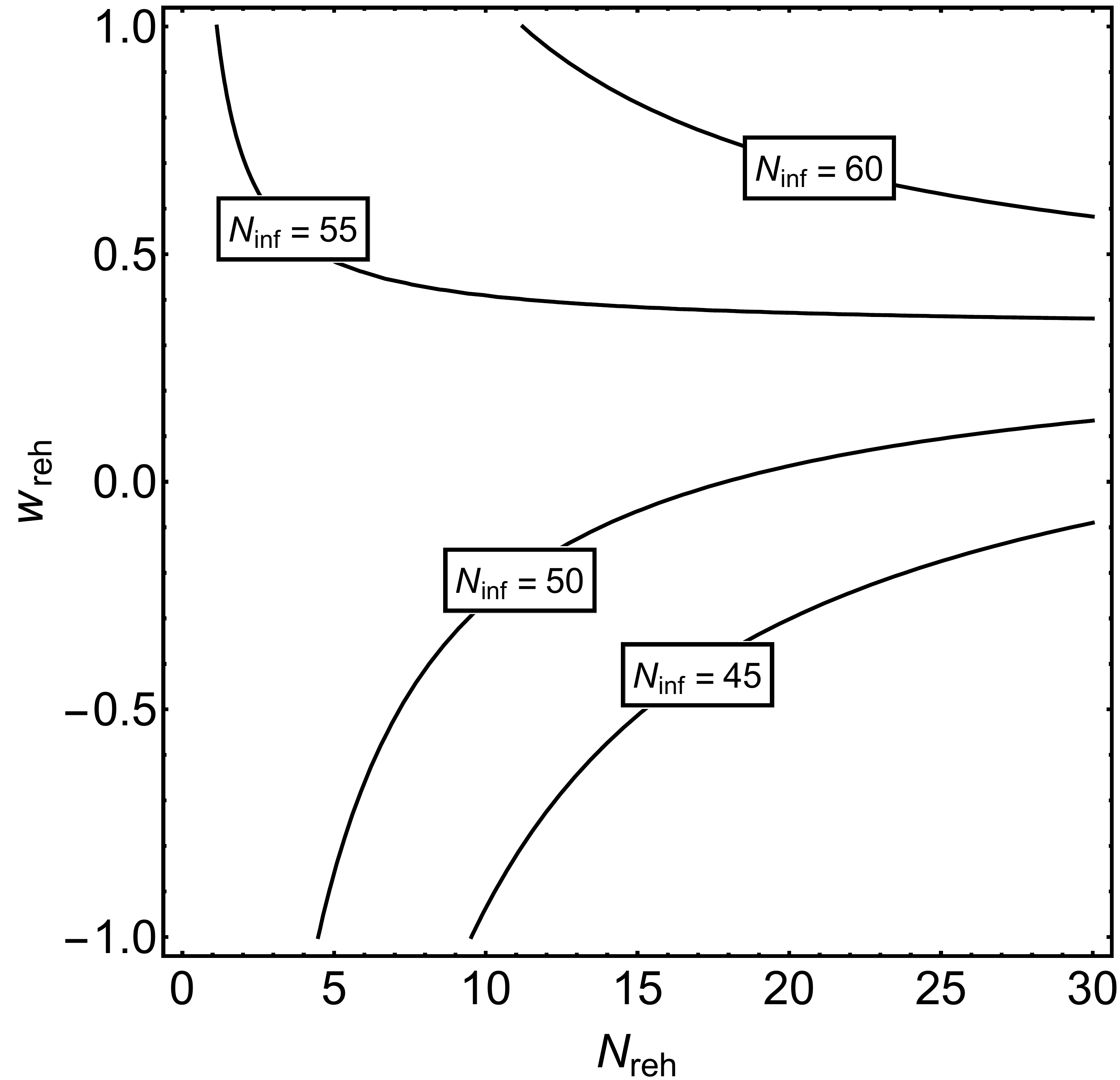}
  \includegraphics[width=7cm]{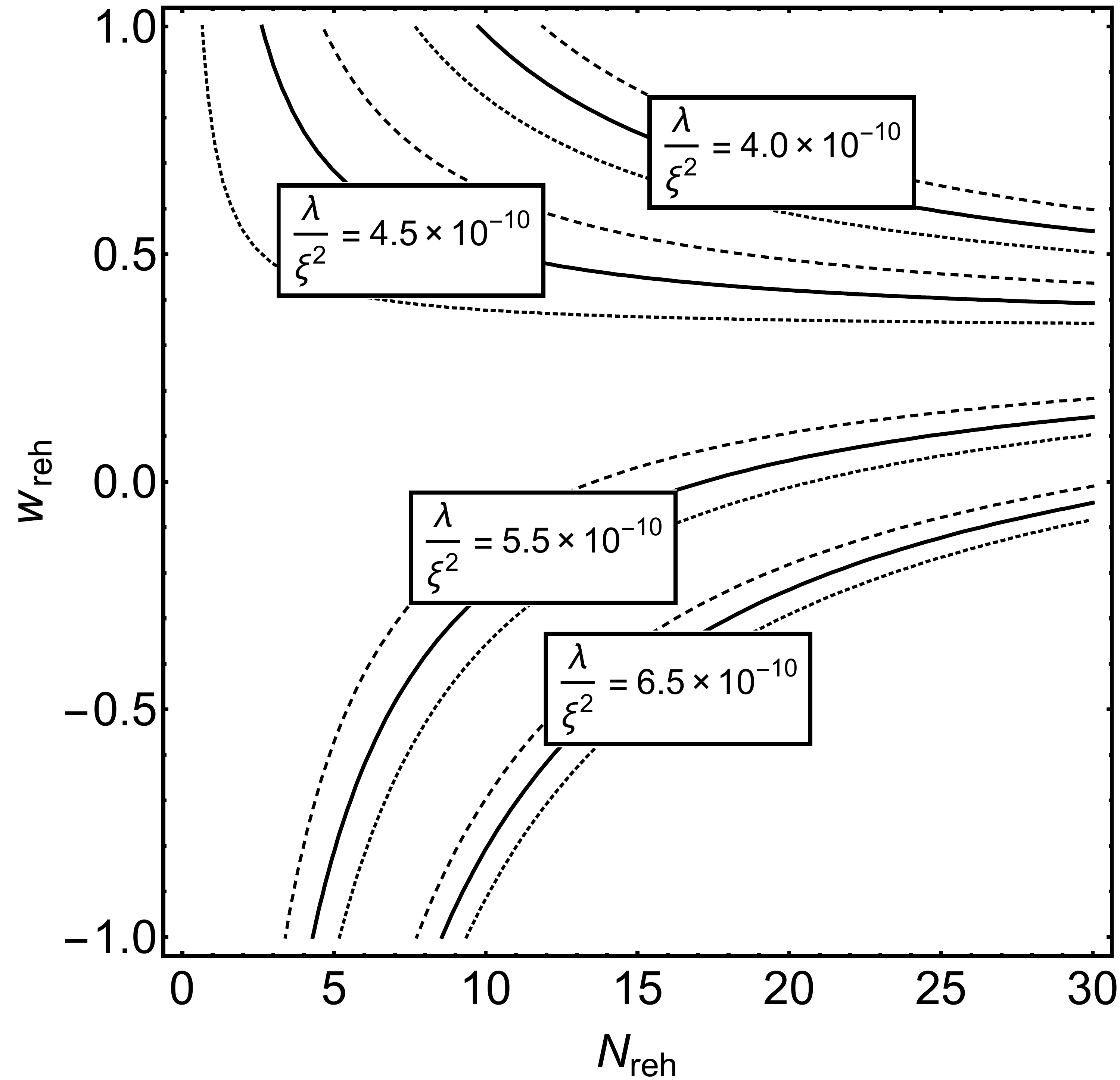}\\
  \caption{Reheating phase diagrams for Higgs inflation. Cosmological predictions of $n_s, r, \alpha_s, Z, T_{\mathrm{reh}}, N_{\mathrm{inf}}$ are drawn with respect to phase variables $N_{\mathrm{reh}}$ and $w_{\mathrm{reh}}$. The dashed contour lines in the last panel are due to different input values of $\ln(10^{10}A_s)=3.094\pm0.034$ from the Planck 2015 normalization~\cite{Ade:2015lrj}. It can be shown in the first panel that almost all possible reheating processes are allowed within the $1\sigma$ region of $n_s=0.9645\pm0.0049$ reported by Planck 2015 TT,TE,EE$+$lowP~\cite{Ade:2015lrj}. This insensitivity of cosmological predictions on the reheating phase can also be seen in other panels.}\label{fig:Higgs}
\end{figure*}

The reheating phase diagrams with respect to $n_s, r, \alpha_s, Z, T_{\mathrm{reh}}, N_{\mathrm{inf}}$ for Higgs inflation can be shown simultaneously in Fig. \ref{fig:Higgs}
with input values of $g_{\mathrm{reh}}=106.75$ and $\ln(10^{10}A_s)=3.094\pm0.034$ from the Planck 2015 normalization~\cite{Ade:2015lrj}. The dashed contour lines in the last panel are due to different input values of $\ln(10^{10}A_s)=3.094\pm0.034$ from the Planck 2015 normalization~\cite{Ade:2015lrj}. However, the dashed contour lines in other panels are too close to the solid contour lines to tell any difference. Therefore, it suffices to take the mean value for $\ln(10^{10}A_s)$ when other inflationary models are concerned. It can be seen in the first panel of Fig. \ref{fig:Higgs} that almost all possible reheating processes are allowed within the $1\sigma$ region of $n_s=0.9645\pm0.0049$ reported by Planck 2015 TT,TE,EE$+$lowP~\cite{Ade:2015lrj}. This insensitivity of cosmological predictions on the reheating phase can be shown in other panels too. Therefore, the cosmological predictions (including reheating temperature as already shown in Ref. \cite{Gong:2015qha} in the $n_s\!-\!r$ plane) of Higgs inflation are insensitive to its reheating processes given the current precision of CMB measurements. However, the reheating processes should be appreciated for the future measurement of $n_s$ with refined precision up to $1\%$ and direct detection of primordial gravitational waves.

\subsection{Reheating phase diagram for other models}\label{subsec:3.2}

The reheating phase diagram can be also built to other single-field slow-roll inflationary models. Among those inflationary models selected by the Planck Collaboration \cite{Ade:2015lrj}, we choose to study power-law potential, hilltop inflation, natural inflation, SB SUSY inflation, and superconformal $\alpha$-attractors E/T-models. Since the models closest to Higgs inflation (equivalent to $R^2$ inflation to the lowest order in slow-roll approximation at tree level) in terms of Bayes evidence are brane inflation and exponential inflation, we expect the outcome of these two models would be essentially the same as Higgs inflation. From now on, we adopt the fiducial value $g_{\mathrm{reh}}=10^3$ and the mean value $\ln(10^{10}A_s)=3.092$ from Planck 2015 TT,TE,EE$+$lowP for $\Lambda$CDM+r, of which $n_s=0.9652^{+0.0093}_{-0.0091}$ and $r<0.106$ with $95\%$ limits at pivot scale $k_*=0.05\,\mathrm{Mpc}^{-1}$ will be used to constrain the reheating phase diagrams. As an aside, the reheating phase diagrams with respect to the $e$-folding number $N_{\mathrm{inf}}$ during inflation are also presented for all models.

\subsubsection{Power-law potential}\label{subsubsec:3.2.1}

\begin{figure*}
  \includegraphics[width=7.8cm]{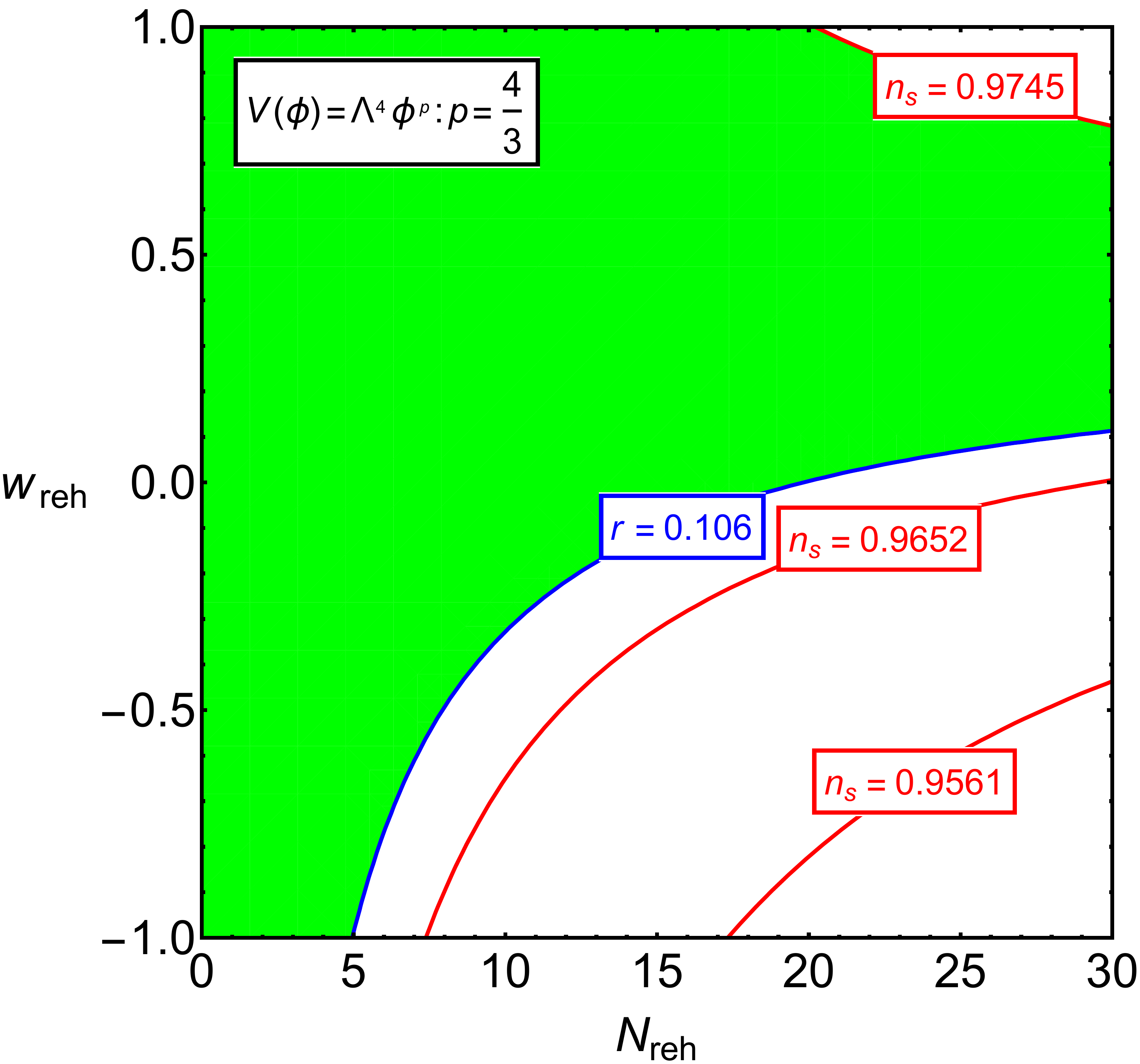}
  \includegraphics[width=7.4cm]{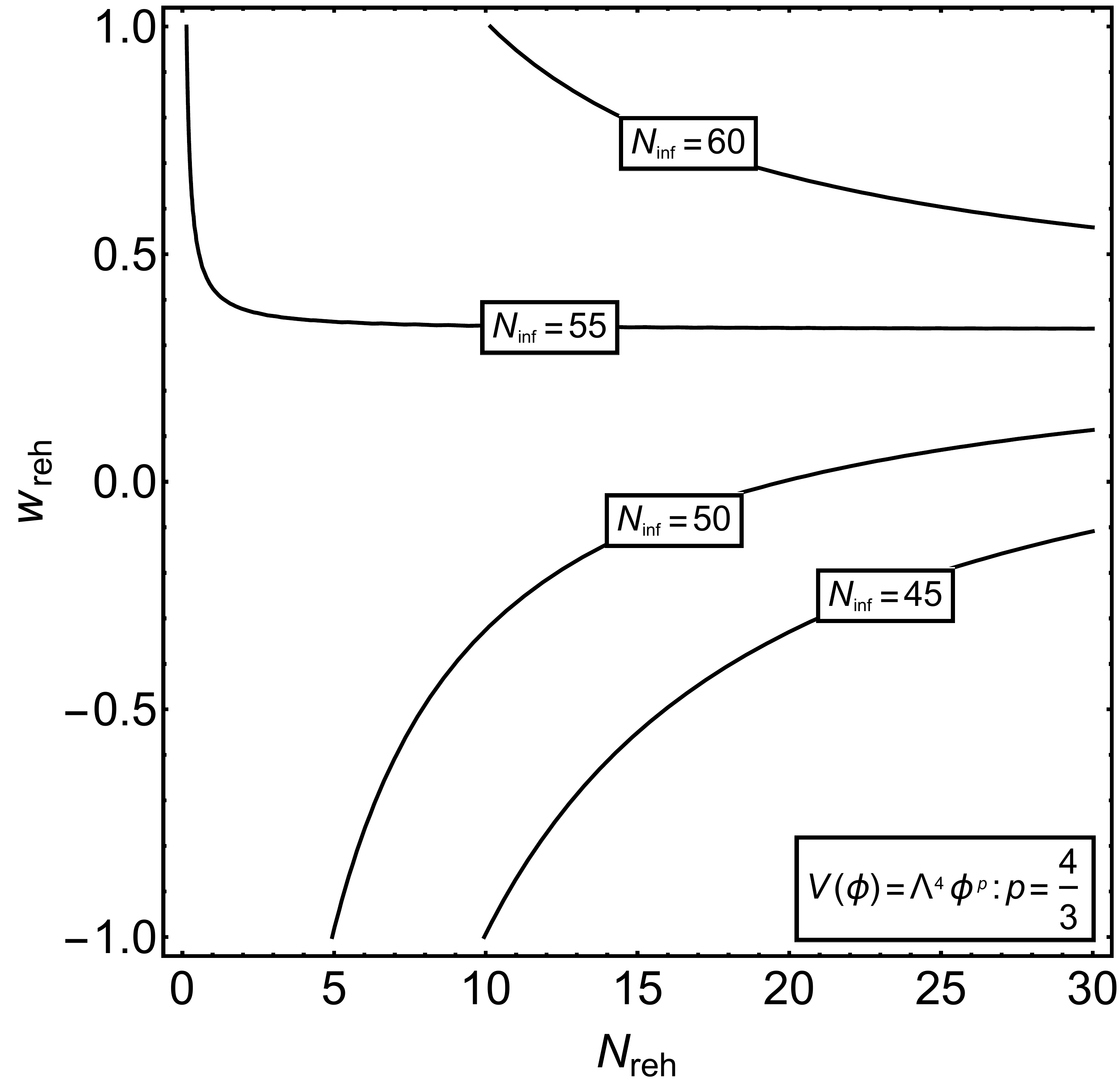}\\
  \includegraphics[width=7.8cm]{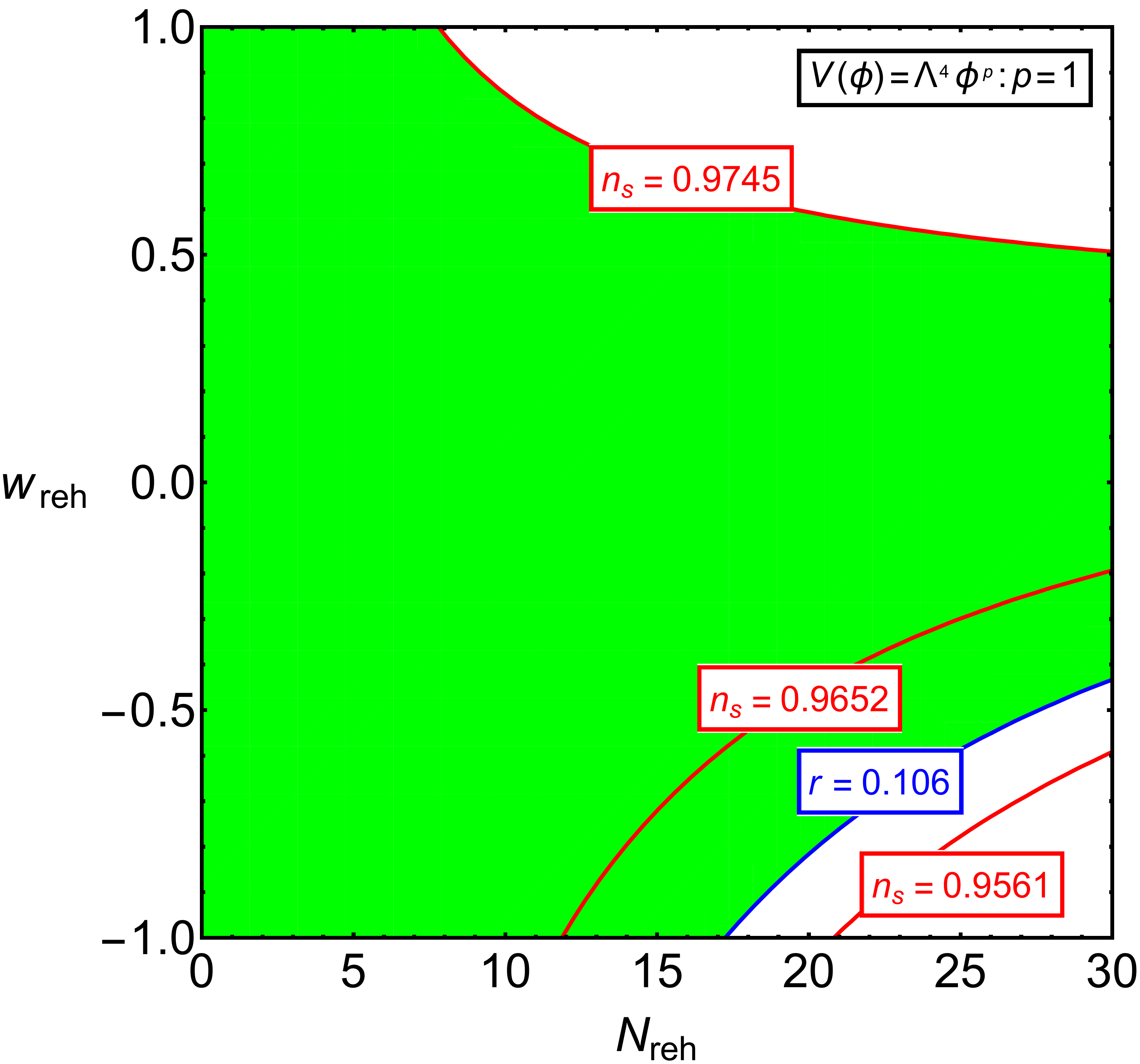}
  \includegraphics[width=7.4cm]{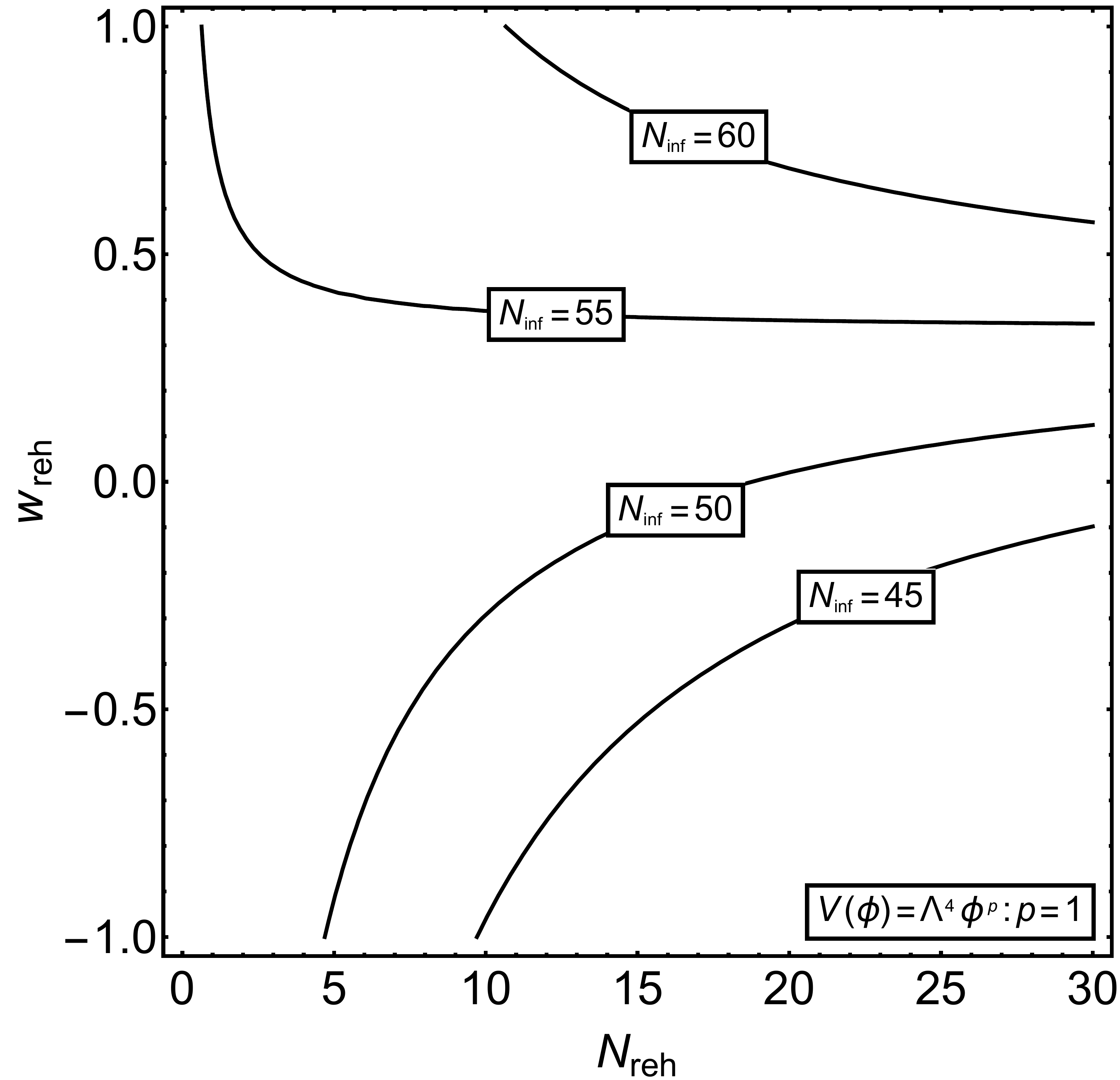}\\
  \includegraphics[width=7.8cm]{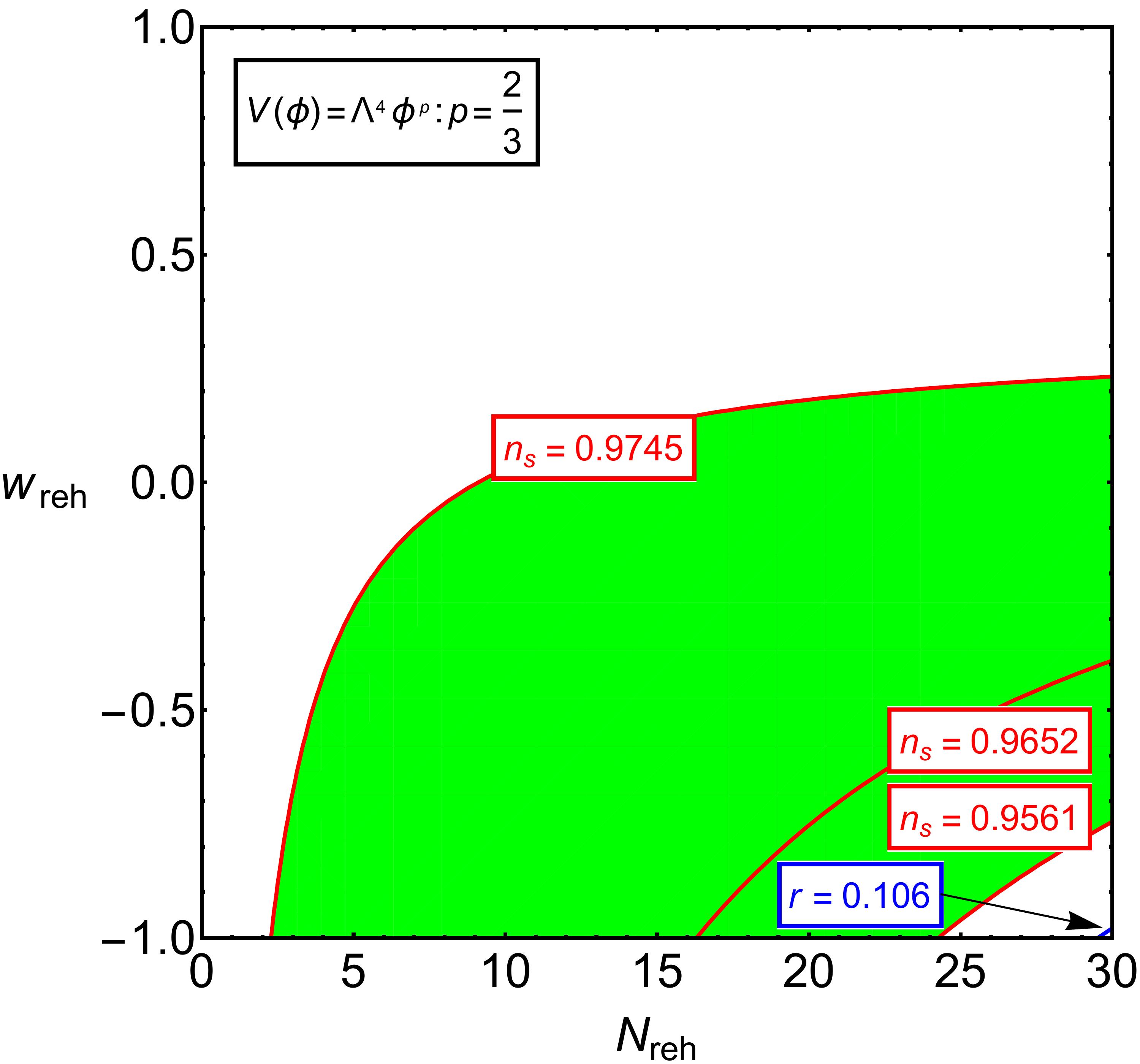}
  \includegraphics[width=7.4cm]{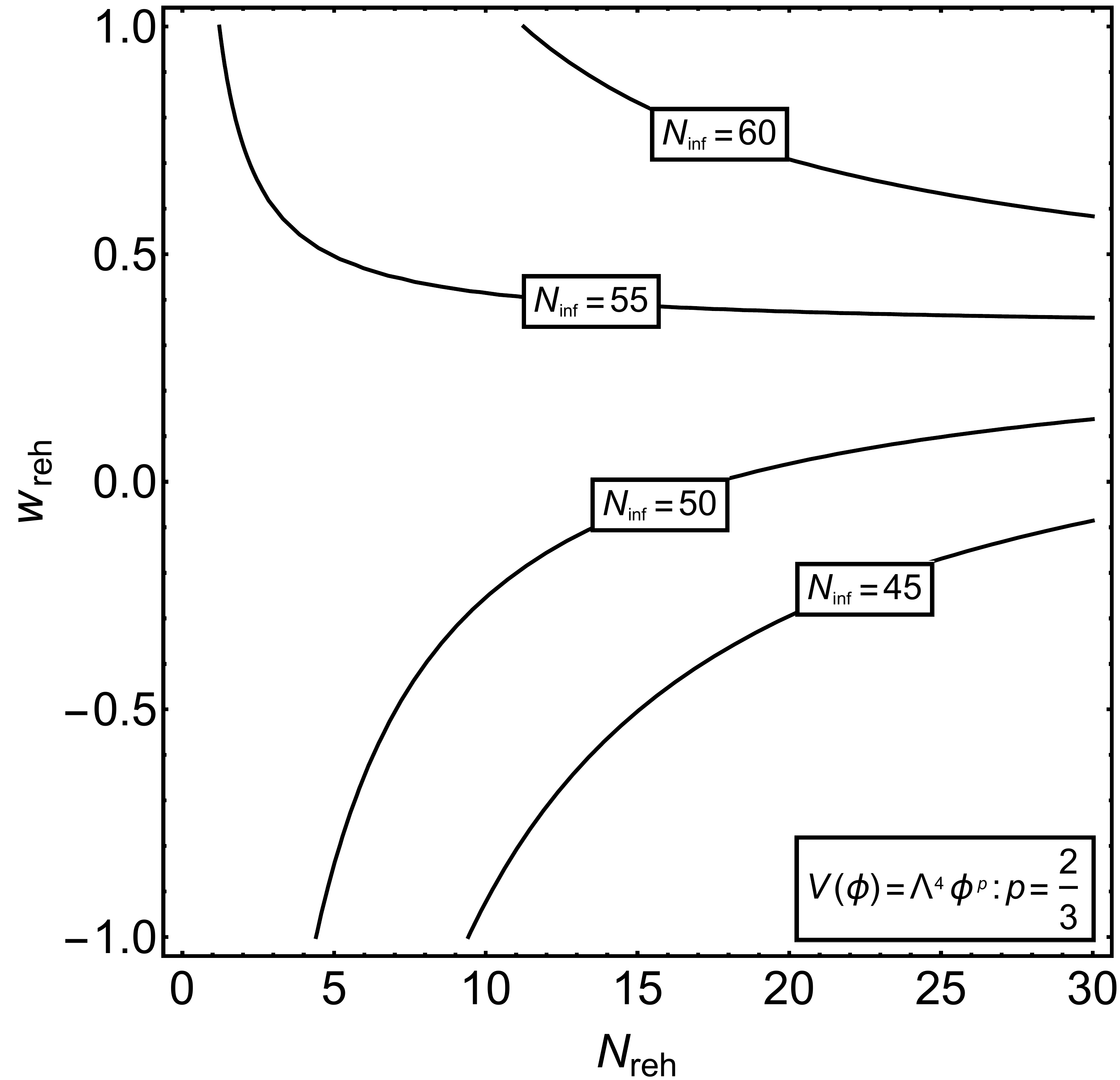}\\
  \caption{Reheating phase diagrams for power-law potential with $p=4/3,1,2/3$ from the top row down. In the left column, the green colored region is specified by requiring requiring $0.9561<n_s<0.9745$ and $r<0.106$. The reheating phase diagrams with respect to $N_{\mathrm{inf}}$ are also presented in the right column. }\label{fig:chaotic}
\end{figure*}

Inflation models with power-law potential~\cite{Linde:1983gd} motivated by axion monodromy~\cite{Silverstein:2008sg,McAllister:2008hb} take values, such as $p=4/3,1,2/3$ for
\begin{equation}
V(\phi)=\Lambda^4\phi^p.
\end{equation}
The slow-roll approximations give
\begin{eqnarray}
\epsilon(\phi)&=&p^2/(2\phi^2),\\
\eta(\phi)&=&p(p-1)/\phi^2,\\
\phi_{\mathrm{end}}&=&p/\sqrt{2},\\
N&=&(\phi_N^2-\phi_{\mathrm{end}}^2)/(2p),
\end{eqnarray}
which are necessary to carry out their reheating phase diagrams in Fig. \ref{fig:chaotic}.
By requiring $0.9561<n_s<0.9745$ and $r<0.106$ according to current constraints on inflation from Planck 2015 TT,TE,EE$+$lowP~\cite{Ade:2015lrj}, one finds the parameter spaces of its reheating phase specified by the green color in the $N_{\mathrm{reh}}\!-\!w_{\mathrm{reh}}$ plane. An interesting observation is that inflation models with a larger green area have larger Bayes factors, as shown in Table 6 in Ref.~\cite{Ade:2015lrj}. It is worth noting that the axion monodromy inflation with $\phi^{2/3}$ potential, which lies at the edge of the $95\%$ confidence region in the $n_s\!-\!r$ plane constrained by Planck 2015 TT,TE,EE$+$lowP, can actually meet the current constraints on inflation from Planck 2015 if the reheating processes~\cite{Moghaddam:2015ava} are taken into account. We did not present here the reheating phase diagrams for power-law potential with $p\geqslant2$, because their reheating phase diagrams simply have no green region at all. (To put it in other words, the regions allowed for $0.9561<n_s<0.9745$ and regions allowed for $r<0.106$ have no intersection.)

\subsubsection{Hilltop inflation}\label{subsubsec:3.2.2}

\begin{figure*}
  \includegraphics[width=7.8cm]{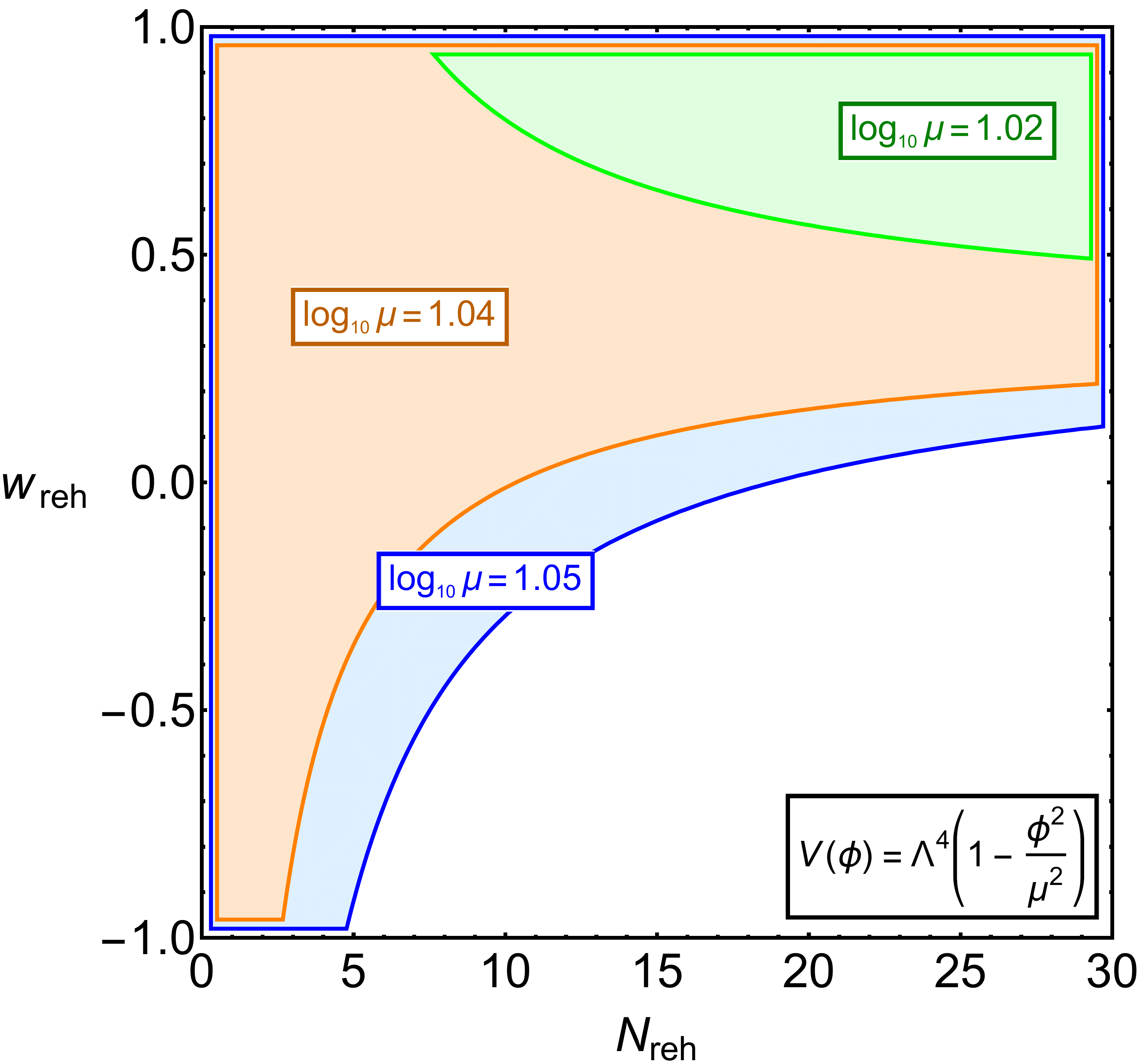}
  \includegraphics[width=7.4cm]{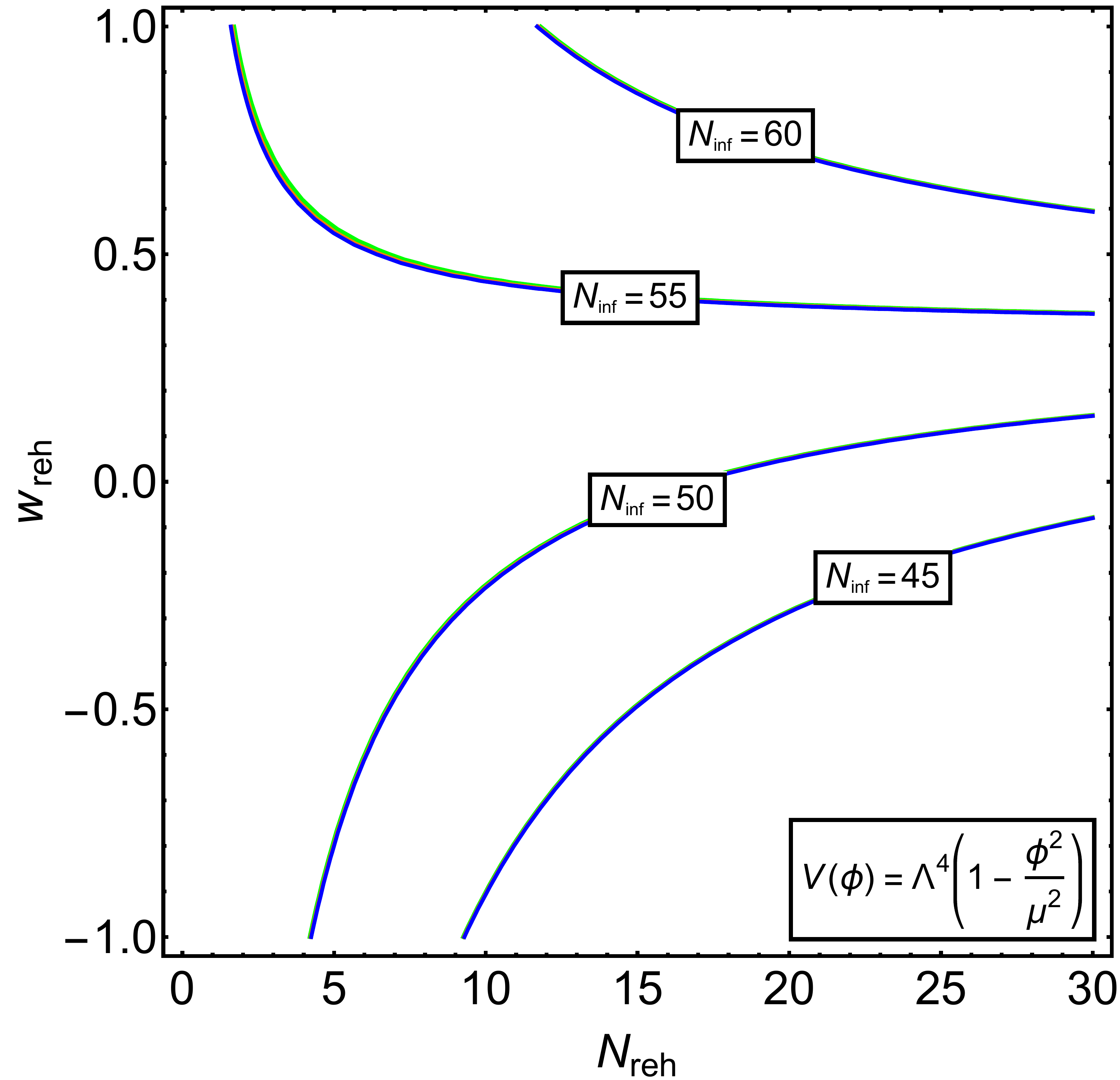}\\
  \includegraphics[width=7.8cm]{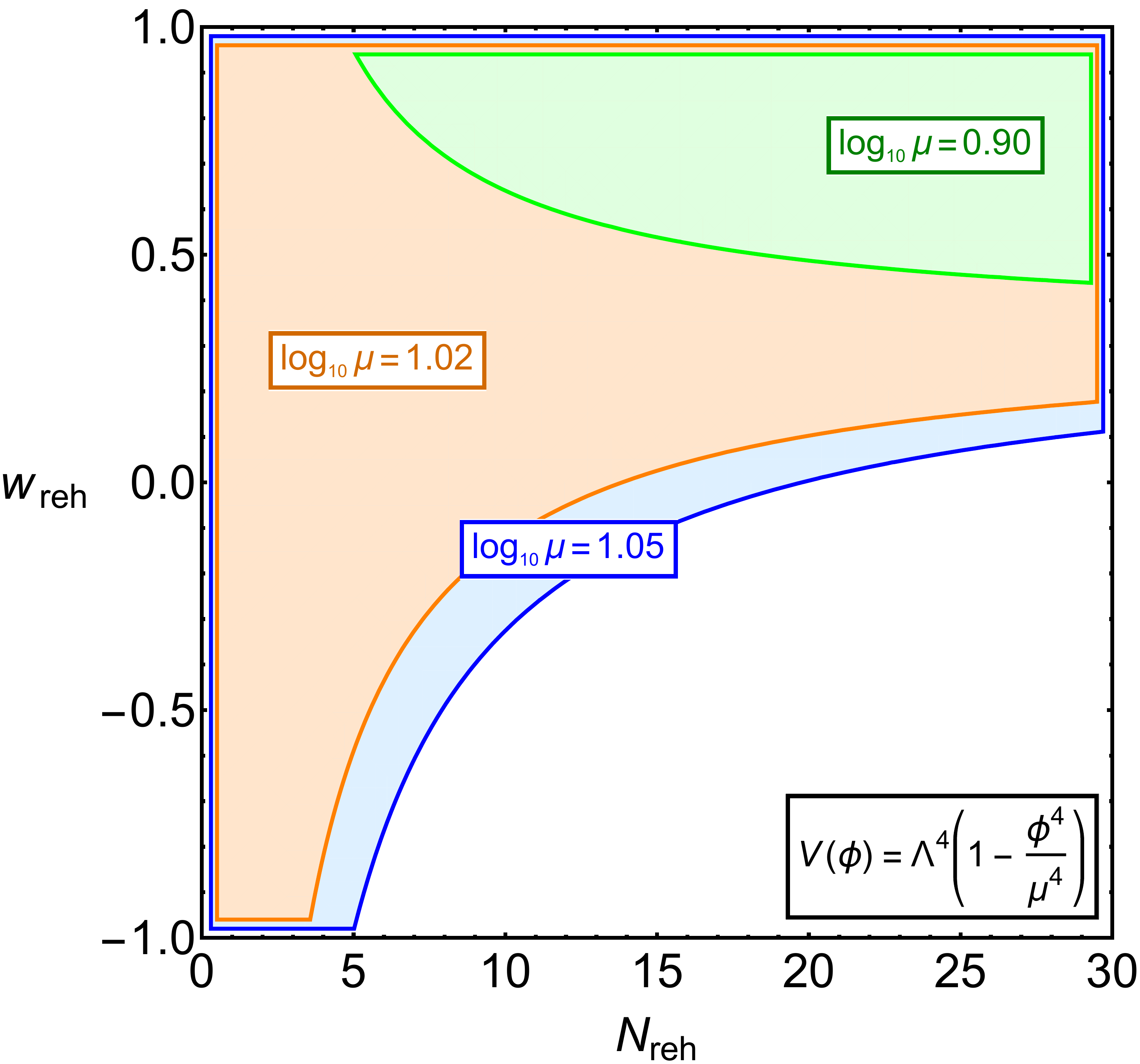}
  \includegraphics[width=7.4cm]{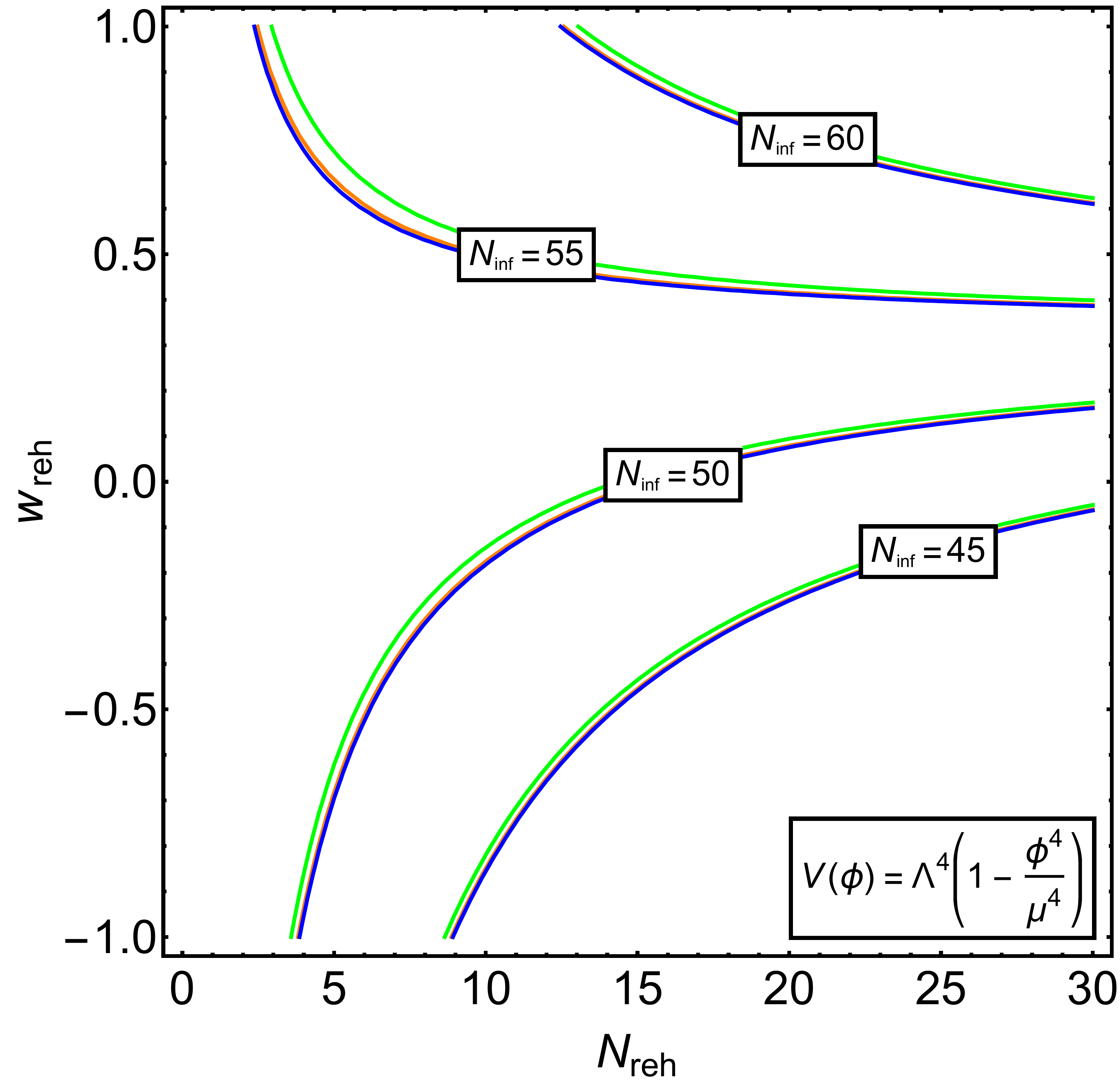}\\
  \caption{Reheating phase diagrams for hilltop inflation with $p=2$ (first row) and $p=4$ (second row). In the left column, each colored region is specified by requiring $0.9561<n_s<0.9745$ and $r<0.106$ with respect to different values of parameter $\mu$. In the right column, reheating phase diagrams with respect to $N_{\mathrm{inf}}$ are presented for the corresponding choices of parameter $\mu$ as in the left column.}\label{fig:hilltop}
\end{figure*}

Hilltop inflation~\cite{Boubekeur:2005zm} with inflationary potential
\begin{equation}
V(\phi)=\Lambda^4\left(1-\frac{\phi^p}{\mu^p}+\cdots\right)
\end{equation}
considered here takes the small field limit $\phi\ll\mu$ with super-Planckian VEV $\mu\gg1$.
The slow-roll approximations give
\begin{eqnarray}
\epsilon_1(\phi)&=&-\frac{\dot{H}}{H^2}=\frac{p^2}{2\mu^2}\frac{(\phi/\mu)^{2p-2}}{(1-(\phi/\mu)^p)^2},\\
\epsilon_2(\phi)&=&\frac{\dot{\epsilon_1}}{H\epsilon_1}=\frac{2p}{\mu^2}\frac{p-1+(\phi/\mu)^p}{(1-(\phi/\mu)^p)^2}(\phi/\mu)^{p-2},\\
\phi_{\mathrm{end}}&=&\mu-1/\sqrt{2}+(p-1)/(4\mu)+\mathcal{O}(1/\mu^2),
\end{eqnarray}
in addition with
\begin{eqnarray}
\nonumber N&=&\frac{\mu^2}{2p}\left[\left(\frac{\phi_N}{\mu}\right)^2-\left(\frac{\phi_{\mathrm{end}}}{\mu}\right)^2\right.\\
&&+\left.\frac{2}{p-2}\left(\left(\frac{\phi_N}{\mu}\right)^{2-p}-\left(\frac{\phi_{\mathrm{end}}}{\mu}\right)^{2-p}\right)\right]
\end{eqnarray}
for $p\neq2$ and
\begin{equation}
N=\frac{\mu^2}{4}\left[\left(\frac{\phi_N}{\mu}\right)^2-\left(\frac{\phi_{\mathrm{end}}}{\mu}\right)^2
-2\ln\left(\frac{\phi_N/\mu}{\phi_{\mathrm{end}}/\mu}\right)\right]
\end{equation}
for $p=2$.
It was constrained at the $95\%$ C.L. that Planck 2015 favors hilltop inflation with $\log_{10}\mu>1.02(1.05)$ for $p=2,w_{\mathrm{reh}}=0$ (allowing $w_{\mathrm{reh}}$ to vary) and $\log_{10}\mu>1.05(1.02)$ for $p=4,w_{\mathrm{reh}}=0$ (allowing $w_{\mathrm{reh}}$ to vary). However,
the reheating phase diagrams for hilltop inflation presented in Fig. \ref{fig:hilltop}
slightly loosen the bound on $\log_{10}\mu$. Each colored region is specified by requiring $0.9561<n_s<0.9745$ and $r<0.106$ with respect to different values of parameter $\mu$. Larger values of $\mu$ will cover larger parts of parameter space in the $N_{\mathrm{reh}}\!-\!w_{\mathrm{reh}}$ plane. However, lower values of $\mu$ would require more exotic reheating processes beyond theoretically reasonable reheating processes with $N_{\mathrm{reh}}\sim\mathcal{O}(1)$ and $w_{\mathrm{reh}}\in[-1/3,1/3]$.

\subsubsection{Natural inflation}\label{subsubsec:3.2.3}

\begin{figure*}
  \includegraphics[width=7.8cm]{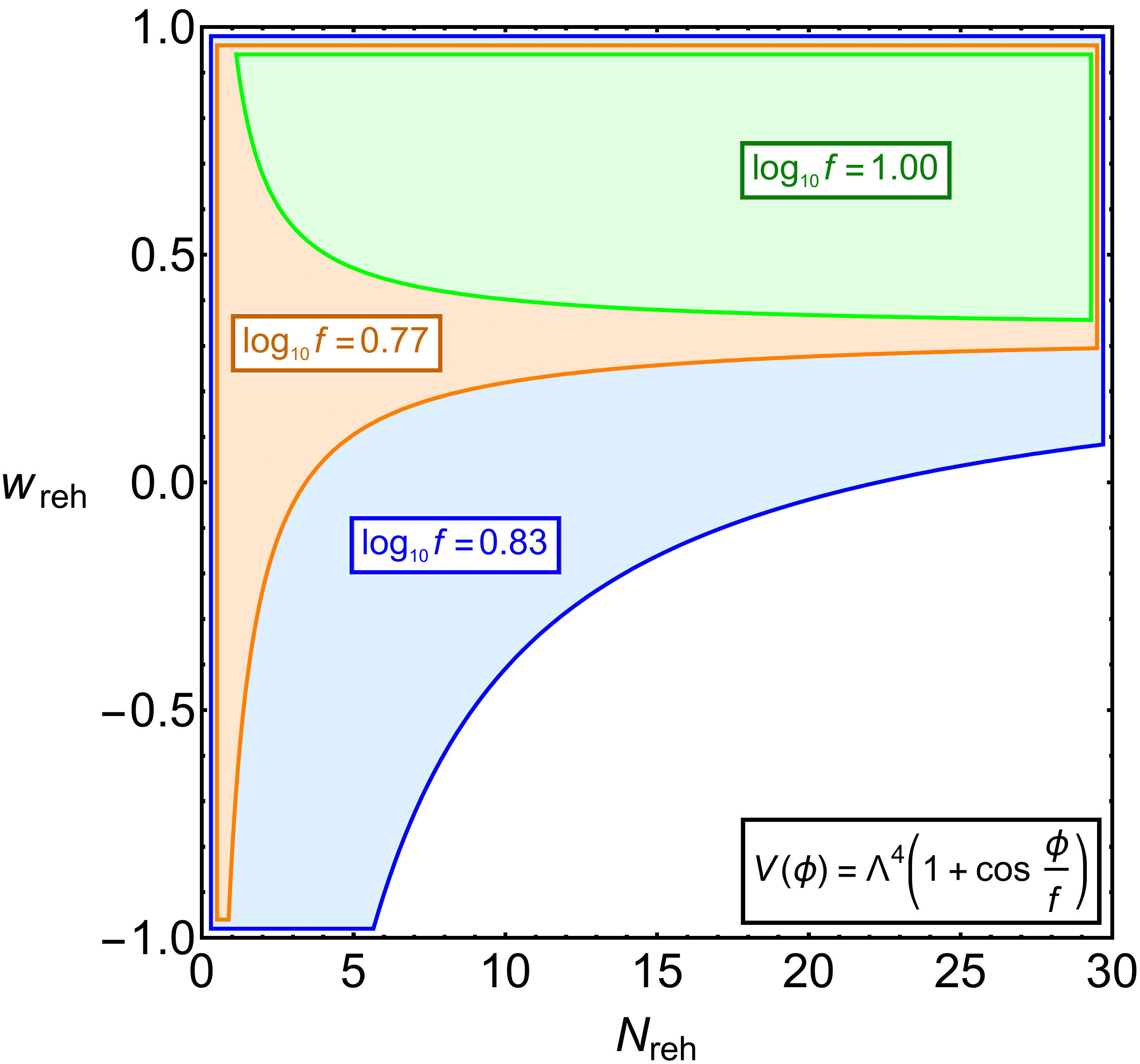}
  \includegraphics[width=7.4cm]{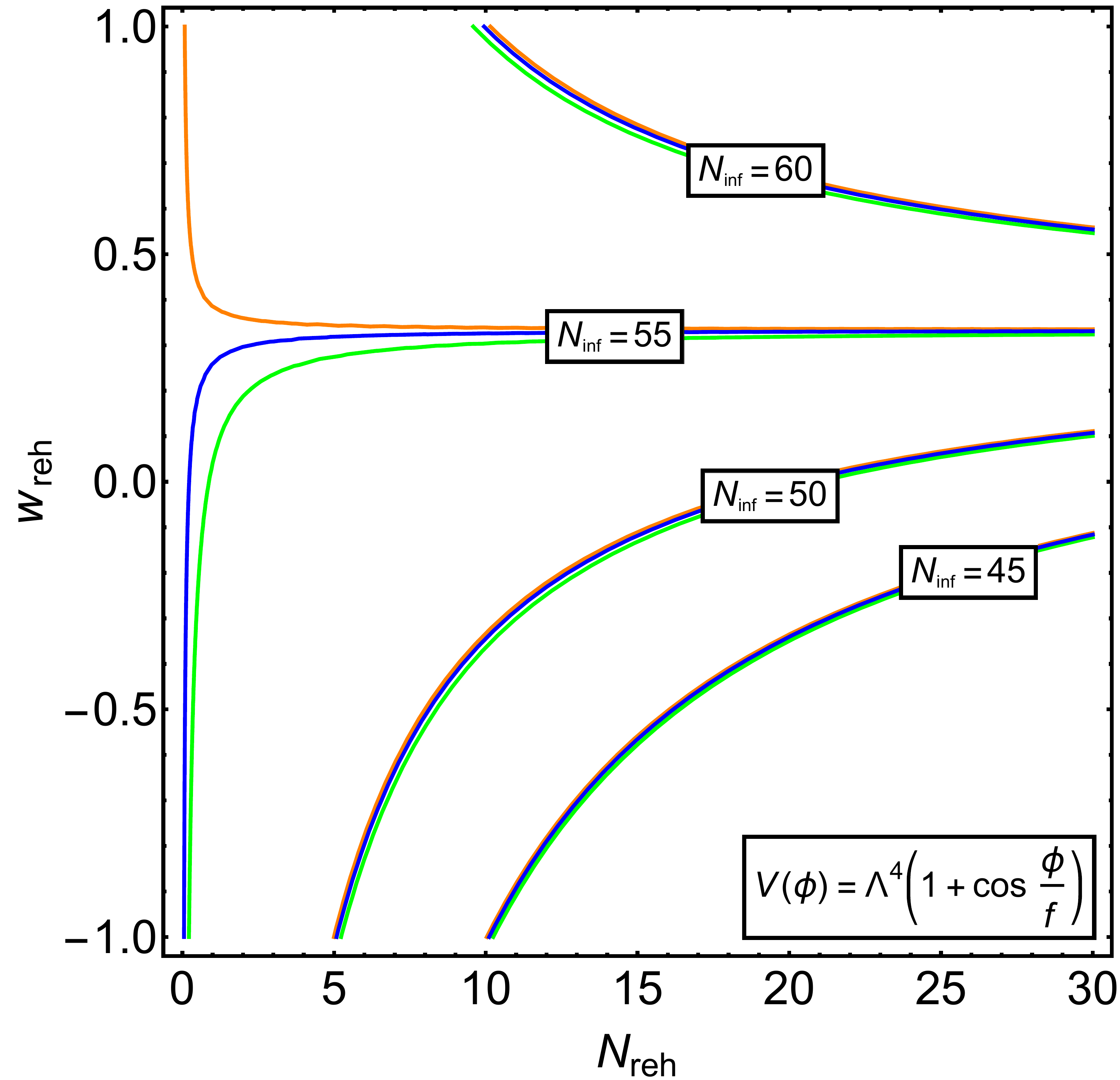}\\
  \caption{Reheating phase diagrams for natural inflation. In the first panel, each colored region is specified by requiring $0.9561<n_s<0.9745$ and $r<0.106$ with respect to different values of parameter $f$. In the second panel, the reheating phase diagram with respect to $N_{\mathrm{inf}}$ is presented for the corresponding choices of parameter $f$ as in the first panel.}\label{fig:natural}
\end{figure*}

Natural inflation \cite{Freese:1990rb,Adams:1992bn} with periodic potential is expressed by
\begin{equation}
V(\phi)=\Lambda^4\left[1+\cos\left(\frac{\phi}{f}\right)\right].
\end{equation}
As far as only $n_s$, $r$ are concerned, natural inflation seems to recover the quadratic chaotic inflation when the scale of curvature of the potential $f\rightarrow\infty$.
The slow-roll approximations give
\begin{eqnarray}
\epsilon_1(\phi)&=&-\frac{\dot{H}}{H^2}=\frac{1}{2f^2}\frac{\sin^2(\phi/f)}{(1+\cos(\phi/f))^2},\\
\epsilon_2(\phi)&=&\frac{\dot{\epsilon_1}}{H\epsilon_1}=\frac{2}{f^2}\frac{1}{1+\cos(\phi/f)},\\
\phi_{\mathrm{end}}&=&f\arccos\left(\frac{1-2f^2}{1+2f^2}\right),\\
N&=&f^2\ln\left(\frac{1-\cos(\phi_{\mathrm{end}}/f)}{1-\cos(\phi_N/f)}\right),
\end{eqnarray}
which are necessary to carry out their reheating phase diagram in Fig. \ref{fig:natural}.
It was constrained at $95\%$ C.L. that Planck 2015 favors natural inflation with $\log_{10}f>0.84(0.83)$ for $w_{\mathrm{reh}}=0$ (allowing $w_{\mathrm{reh}}$ to vary). As in the case of hilltop inflation, the reheating phase diagram presented in Fig. \ref{fig:natural} slightly loosens the bound on $\log_{10}f$, and smaller or larger values of $f$ will require more exotic reheating processes. However, since there are temporarily no observational constraints on reheating phase variables, one cannot simply rule out these parameter spaces.

\subsubsection{Spontaneously broken SUSY}\label{subsubsec:3.2.4}

\begin{figure*}
  \includegraphics[width=7.8cm]{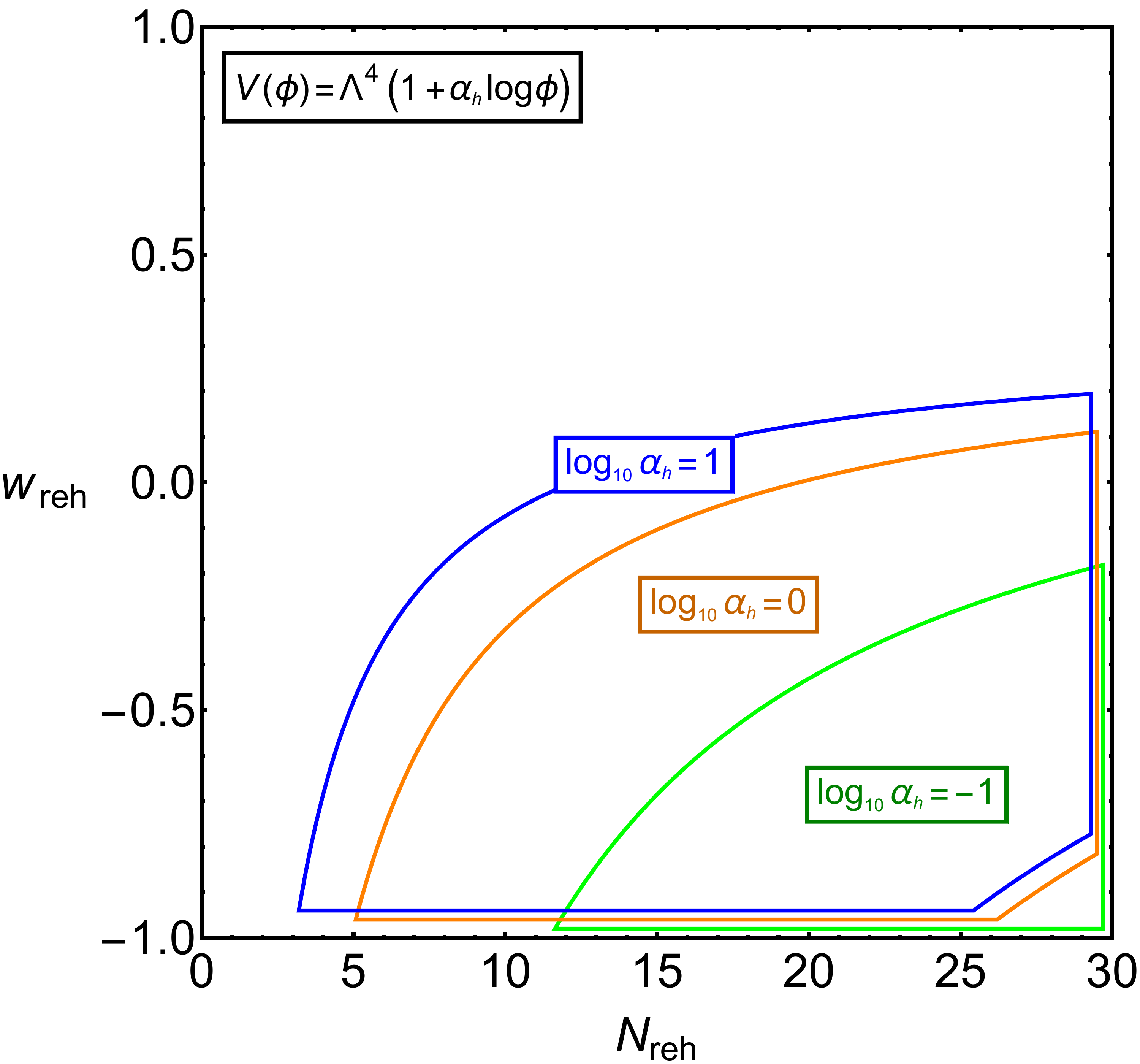}
  \includegraphics[width=7.4cm]{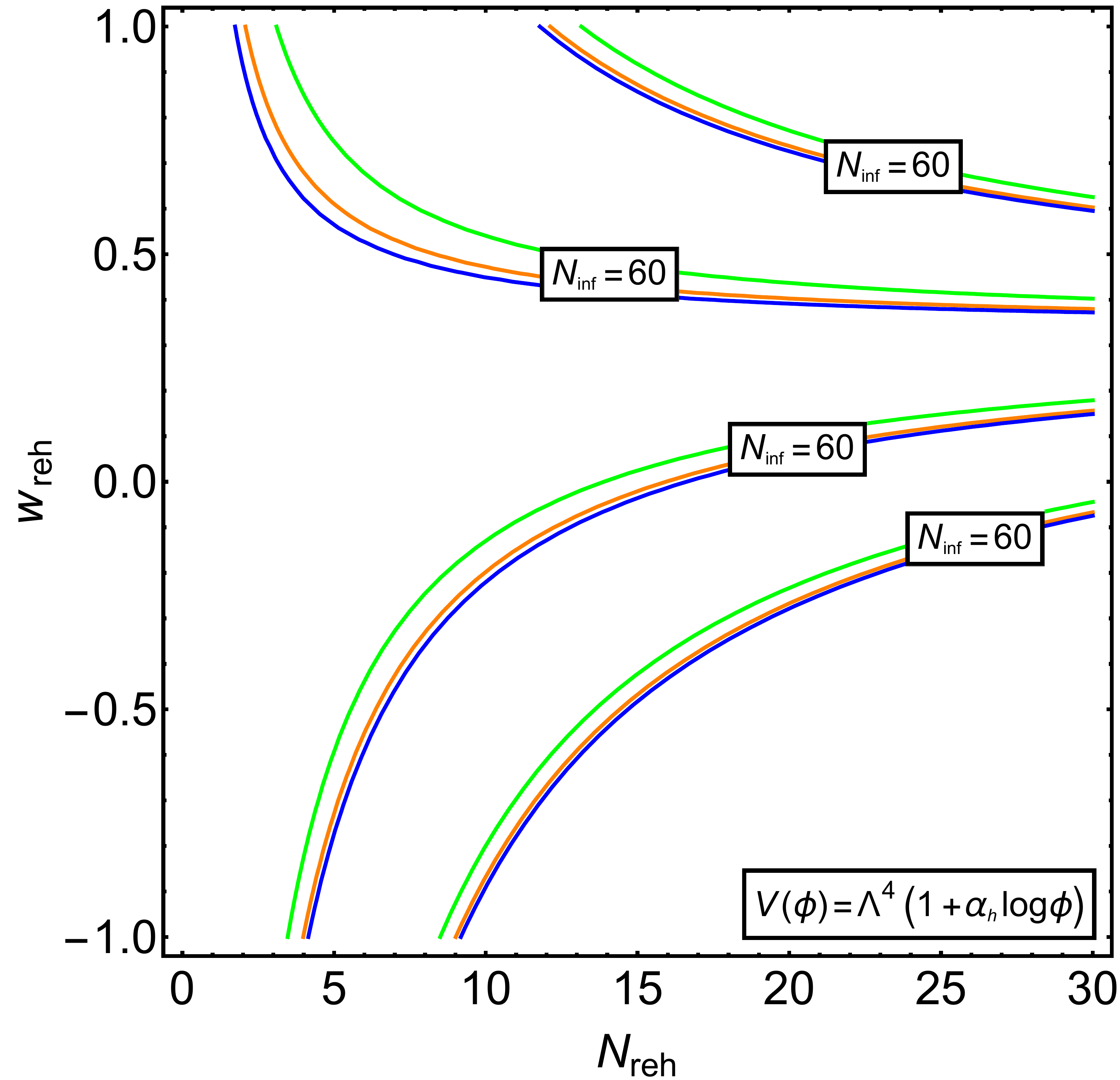}\\
  \caption{Reheating phase diagrams for SB SUSY inflation. In the first panel, each colored region is specified by requiring $0.9561<n_s<0.9745$ and $r<0.106$ with respect to different values of parameter $\alpha_h$. In the second panel, the reheating phase diagram with respect to $N_{\mathrm{inf}}$ is presented for the corresponding choices of parameter $\alpha_h$ as in the first panel.}\label{fig:SUSY}
\end{figure*}

Spontaneously broken SUSY inflation~\cite{Dvali:1994ms} is described by the potential
\begin{equation}
V(\phi)=\Lambda^4(1+\alpha_h\log\phi)
\end{equation}
with a flat prior $[-2.5,1]$ for $\log_{10}\alpha_h$. The slow-roll approximations give
\begin{eqnarray}
\epsilon(\phi)&=&\frac{\alpha_h^2}{2\phi^2(1+\alpha_h\log\phi)^2},\\
\eta(\phi)&=&-\frac{\alpha_h}{\phi^2(1+\alpha_h\log\phi)},\\
\phi_{\mathrm{end}}&=&\sqrt{2}/\sqrt{W(2\exp[{2/\alpha_h}])},\\
\nonumber N&=&\left(\frac{1}{2\alpha_h}-\frac{1}{4}\right)\left(\phi_N^2-\phi_{\mathrm{end}}^2\right)\\
 & &+\frac{1}{4}\left(\phi_N^2\log\phi_N^2-\phi_{\mathrm{end}}^2\log\phi_{\mathrm{end}}^2\right).
\end{eqnarray}
Here $W(z)$ is the Lambert function by definition $z=W(z)\exp[W(z)]$. The reheating phase diagram for SB SUSY inflation is presented in Fig. \ref{fig:SUSY}.
As in the case of natural inflation, smaller values of $\alpha_h$ are allowed if one invokes more exotic reheating processes. Although SB SUSY inflation lies outside the $95\%$ confidence region in the $n_s\!\!-\!r$ plane constrained by Planck 2015 TT,TE,EE$+$lowP, there exist parameter spaces of reheating phase in the $N_{\mathrm{reh}}\!-\!w_{\mathrm{reh}}$ plane to accommodate the Planck constraints on inflation. Therefore, SB SUSY inflation cannot be simply ruled out if the reheating phase is taken into account.

\subsubsection{$\alpha$-attractors}\label{subsubsec:3.2.5}

\begin{figure*}
  \includegraphics[width=7.8cm]{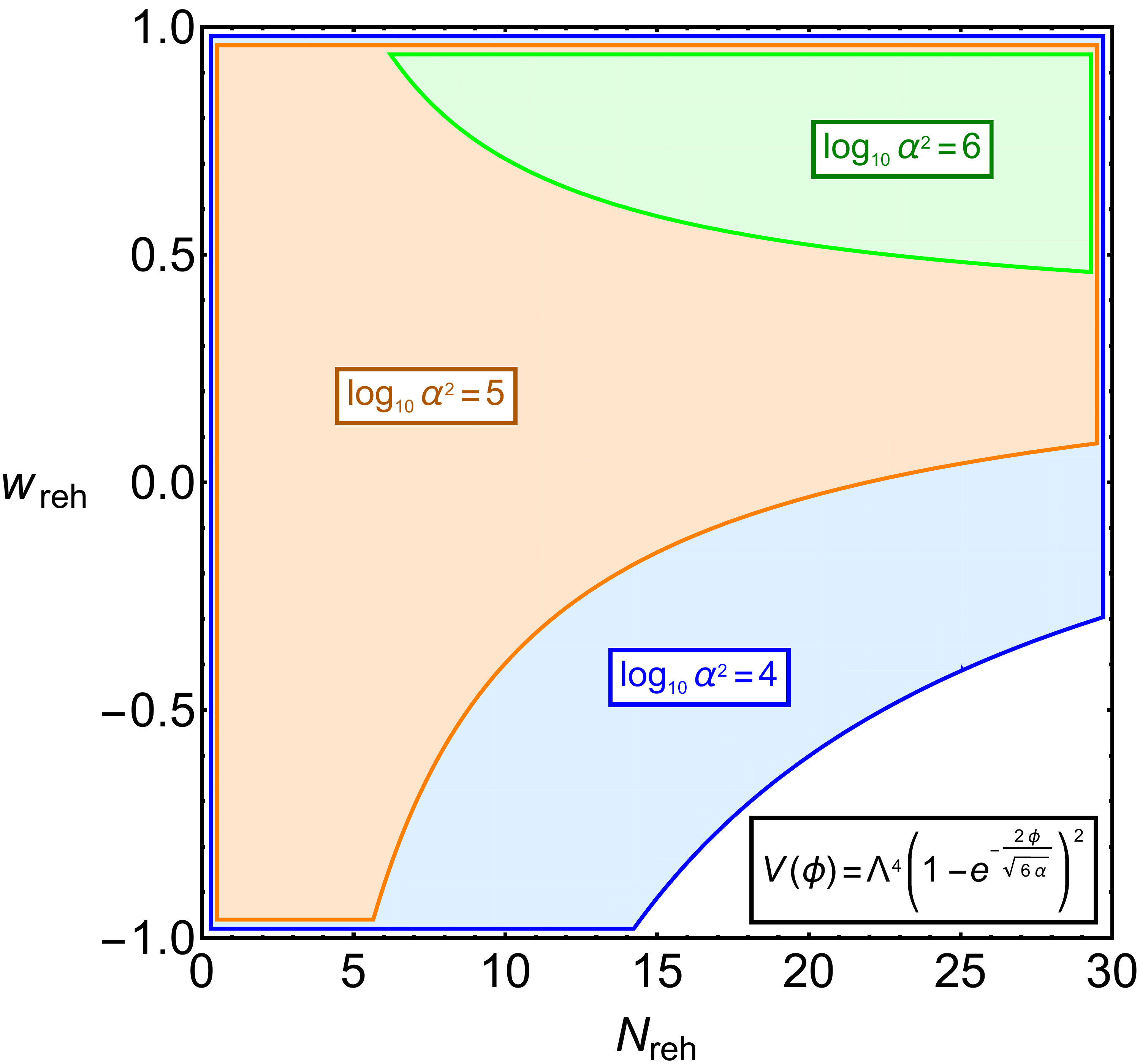}
  \includegraphics[width=7.4cm]{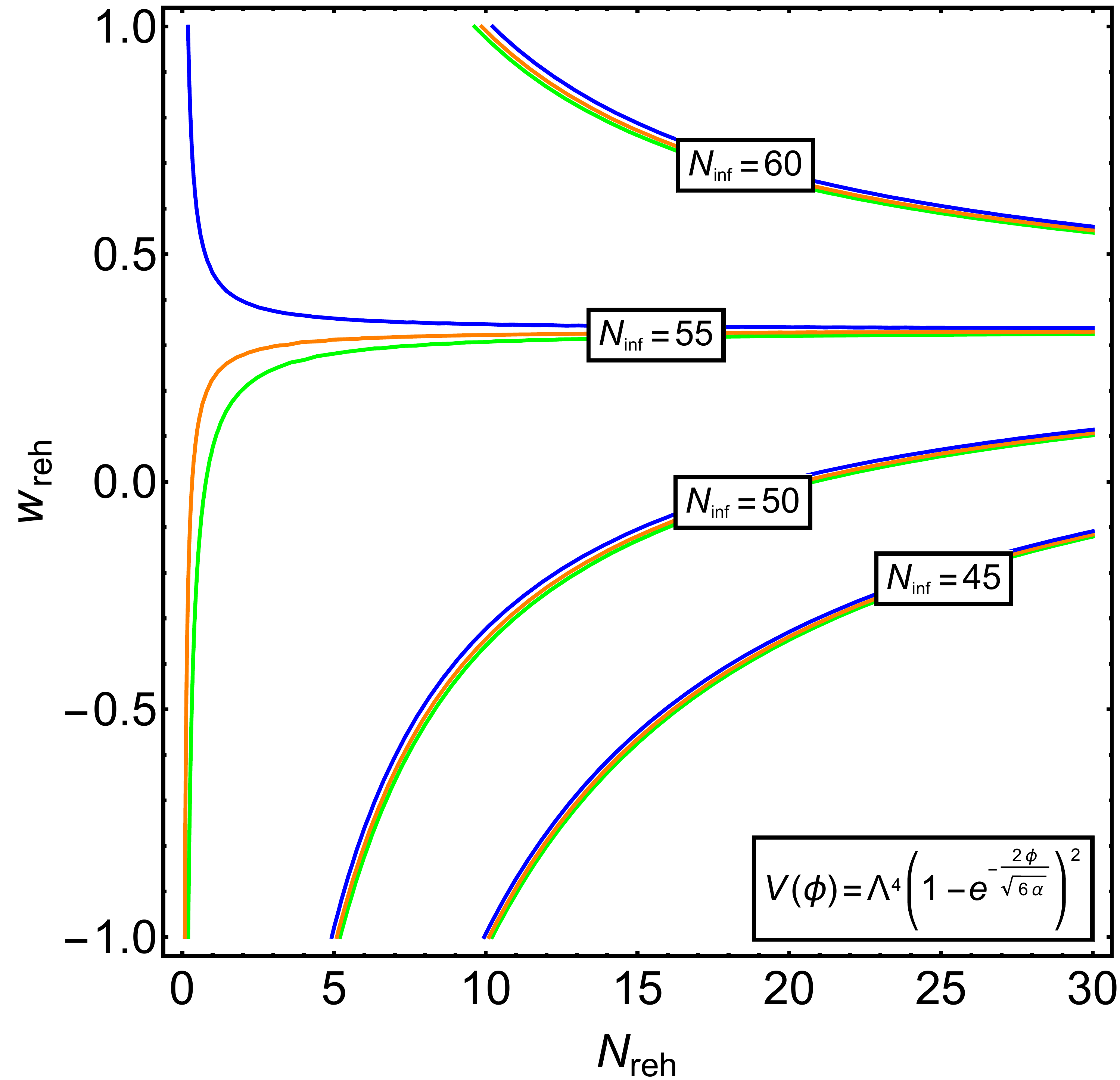}\\
  \includegraphics[width=7.8cm]{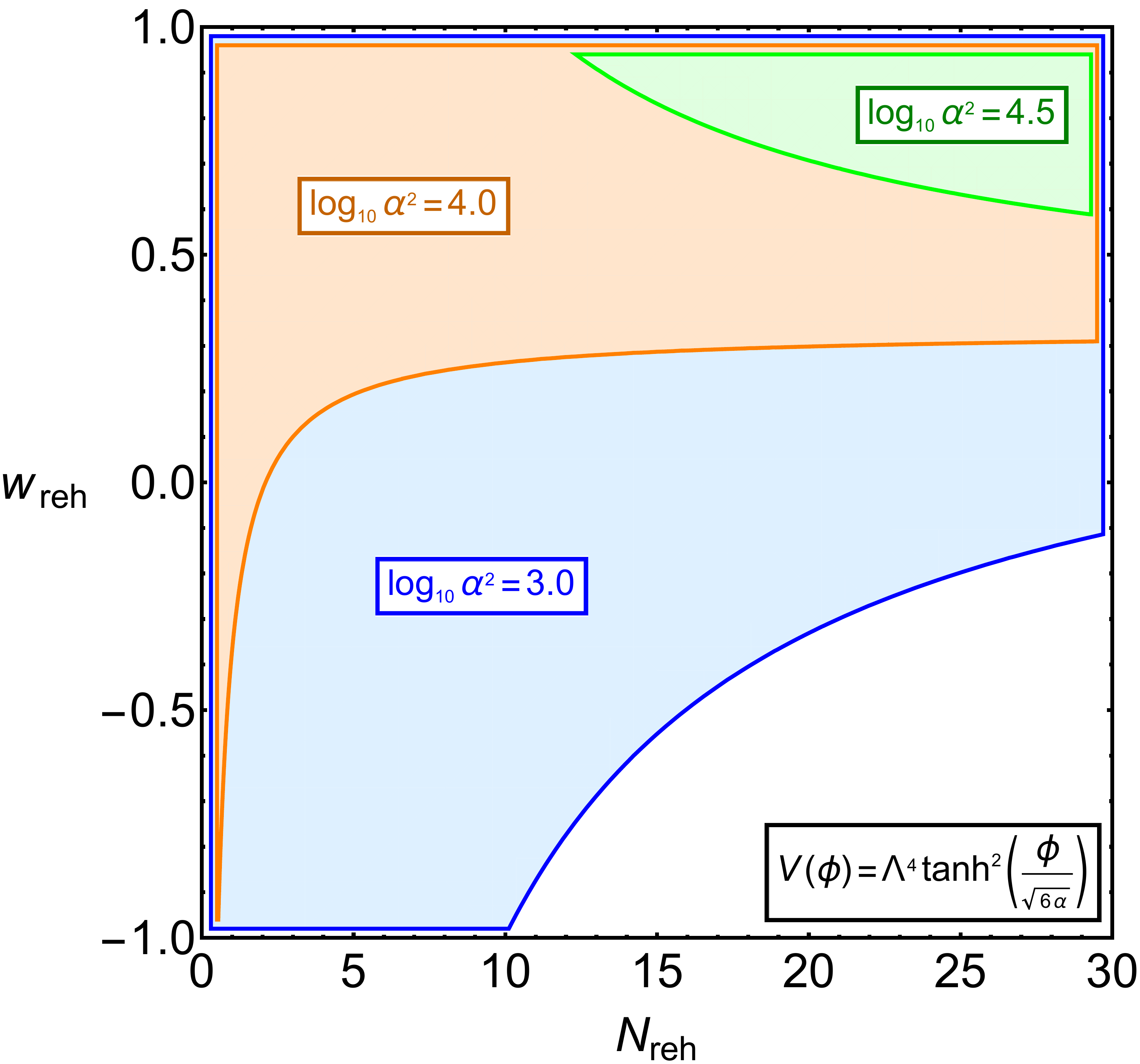}
  \includegraphics[width=7.4cm]{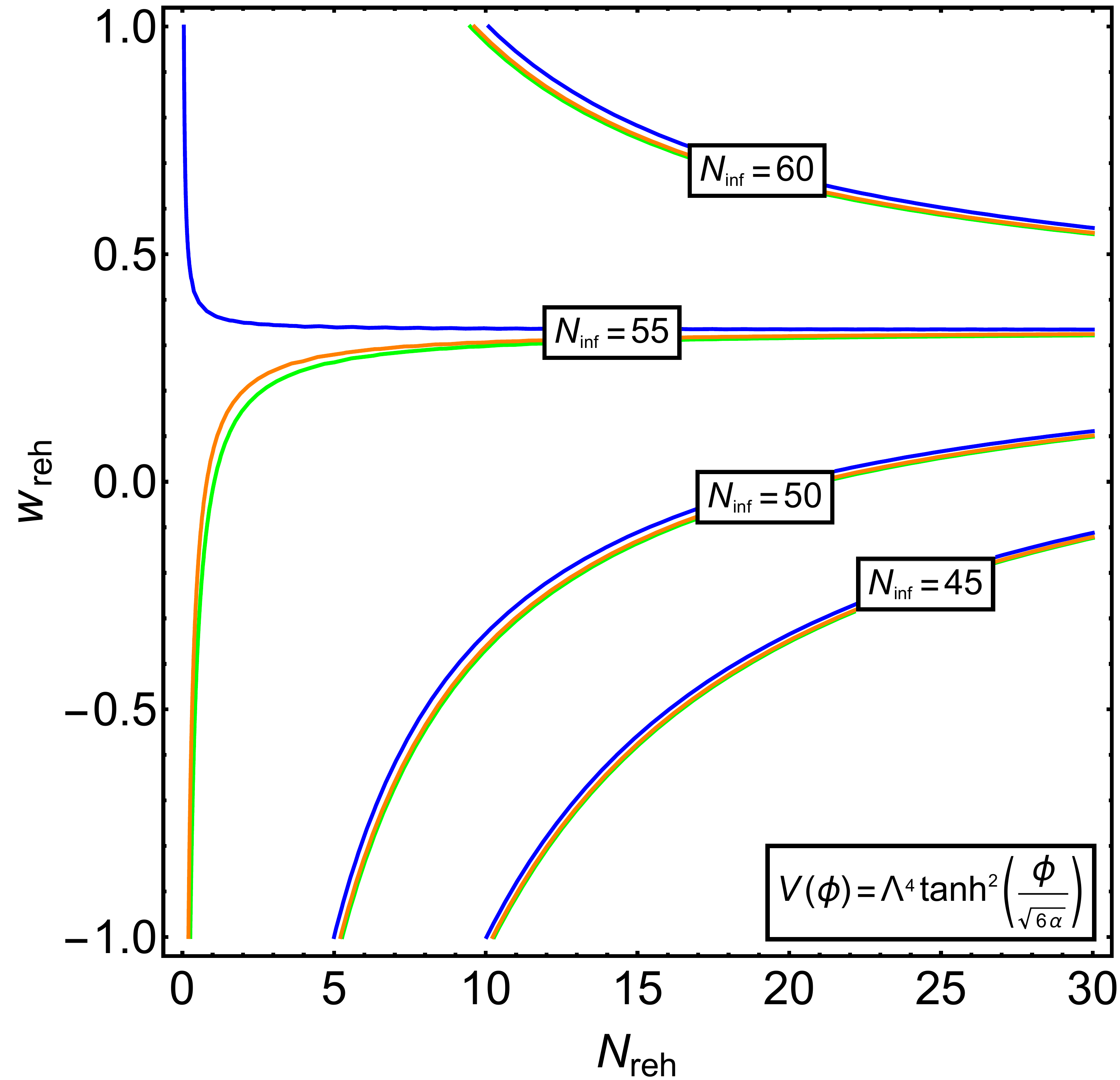}\\
  \caption{Reheating phase diagrams for $\alpha$-attractors E-model (first row) and T-model (second row) with $m=1$. In the left column, each colored region is specified by requiring $0.9561<n_s<0.9745$ and $r<0.106$ with respect to different values of parameter $\alpha$. In the right column, reheating phase diagrams with respect to $N_{\mathrm{inf}}$ are presented for the corresponding choices of parameter $\alpha$ as in the left column.}\label{fig:alpha}
\end{figure*}

$\alpha$-attractors E-models \cite{Kallosh:2013yoa} with exponentially flat potential
\begin{equation}
V(\phi)=\Lambda^4\left(1-e^{-\frac{2\phi}{\sqrt{6\alpha}}}\right)^2
\end{equation}
approach the predictions on $(n_s,r)$ of quadratic inflation for $\alpha\rightarrow\infty$ and the Starobinsky model $(n_s=1-2/N,r=12/N^2)$ for $\alpha=1$ and $\alpha$ attractors $(n_s=1-2/N,r=0)$ for $\alpha\rightarrow0$. Planck 2015 favors $\alpha$-attractors E-models with $\log_{10}\alpha^2<1.7(2.0)$ for $w_{\mathrm{reh}}=0$ (allowing $w_{\mathrm{reh}}$ to vary). However, the reheating phase diagram presented in Fig. \ref{fig:alpha}
slightly loosens the bound on $\alpha$ as expected.

$\alpha$-attractors T-models \cite{Kallosh:2013yoa} with inflationary potential
\begin{equation}
V(\phi)=\Lambda^4\tanh^{2m}\left(\frac{\phi}{\sqrt{6\alpha}}\right)
\end{equation}
approach the predictions on $(n_s,r)$ of power-law potential $\phi^{2m}$ for $\alpha\rightarrow\infty$ and  $\alpha$ attractors  $(n_s=1-2/N,r=0)$ for $\alpha\rightarrow0$. Planck 2015 favors $\alpha$-attractors T-models with $\log_{10}\alpha^2<2.3(2.5)$ for $m=1,w_{\mathrm{reh}}=0$ (allowing $w_{\mathrm{reh}}$ to vary) and $0.2<m<1(m<1)$ for $m\neq1,w_{\mathrm{reh}}=0$ (allowing $w_{\mathrm{reh}}$ to vary). However, the reheating phase diagram presented in Fig. \ref{fig:alpha} slightly loosens the bound on $\alpha$ as expected.

\section{Conclusions}\label{sec:4}

In the $n_s\!-\!r$ plane, one usually characterizes the uncertainties from the reheating phase by the choice of freedom on the $e$-folding number of inflation. In this paper, we characterize the reheating phase by only two effective parameters, $N_{\mathrm{reh}}$ and $w_{\mathrm{reh}}$. Thanks to the fact that observable quantities are insensitive to the effective number of degrees of freedom at the end of reheating phase, we are able to express all other inflationary observables in terms of the phase variables $N_{\mathrm{reh}}$ and $w_{\mathrm{reh}}$. Therefore, for the first time we are able to constrain the parameter space of the reheating phase in the $N_{\mathrm{reh}}\!-\!w_{\mathrm{reh}}$ plane with respect to the constraints on inflation from Planck 2015. For Higgs inflation, the parameter space of the reheating phase covers almost all the $N_{\mathrm{reh}}\!-\!w_{\mathrm{reh}}$ plane, indicating that inflationary predictions of Higgs inflation are insensitive to its reheating processes given the current precision of CMB measurements. However, future refined measurements on the scalar spectral index and direct detection of primordial gravitational waves will constrain the reheating phase variables. For other inflationary models selected by the Planck Collaboration, the constrained parameter spaces of the reheating phase generally loosen the bound on the potential parameters if more exotic reheating processes are allowed. Inflationary models with larger parameter spaces in the reheating phase diagrams generally appear with larger Bayes factors. Since there are only theoretical considerations, not observational constraints on the possible reheating phase, one cannot simply rule out those parameter spaces of the reheating phase with exotic reheating processes even if they lie outside the $95\%$ confidence region in the $n_s\!-\!r$ plane constrained by Planck 2015 TT,TE,EE$+$lowP. Only those inflationary models with no allowed parameter space of the reheating phase in the $N_{\mathrm{reh}}\!-\!w_{\mathrm{reh}}$ plane can be certainly ruled out.

It should be acknowledged that in the case of warm inflation scenarios~\cite{Berera:1995ie,Berera:1995wh}, a separate reheating phase is not necessary ~\footnote{We thank an anonymous referee for pointing this out to us.}.  Unlike the standard scenarios, where the inflaton field has to be coupled to other degrees of freedom in order to transfer its vacuum energy to reheat the Universe until the radiation era finally takes over, the inflaton field could slowly dissipate its kinetic energy into radiation; thus, the relative abundance of radiation may slowly increase during the inflation phase until it smoothly takes over. When taking these dissipative effects into consideration, one can actually accommodate, for example, the $\lambda\phi^4$ model, with the Planck data for a nearly thermalized state in a supersymmetric realization of warm inflation with renormalizable interactions~\cite{Bartrum:2013fia}.

\begin{acknowledgments}
We would like to thank Peng-Xu Jiang and Jian-Wei Hu for helpful discussions.
R.G.C. is supported by the Strategic Priority Research Program of the Chinese Academy of Sciences, Grant No. XDB09000000.
Z.K.G. is supported by the National Natural Science Foundation of China, Grants No. 11175225 and No. 11335012.

\paragraph*{Note added} Reference~\cite{Gong:2015qha} recently showed up on arXiv. We both follow the same method to study a similar problem for Higgs inflation but with different angles. We characterize the reheating phase with $N_{\mathrm{reh}}$ and $w_{\mathrm{reh}}$ and express every other observable in terms of these two phase variables. The impact from various reheating processes on the inflationary predictions can be shown with respect to not only the reheating temperature but also other cosmological observables. We further study other inflationary models selected by the Planck Collaboration and find that the reheating phase diagram can be used to constrain the parameter space of the reheating phase to meet current constraints on inflation.
\end{acknowledgments}

\bibliographystyle{apsrev4-1}
\bibliography{ref}

\end{document}